\documentclass{emulateapj}
\usepackage{natbib,float,longtable,epsfig,amsfonts,amsmath,graphicx,apjfonts}
\def\swift{{\it Swift}}
\def\chandra{{\it Chandra}}
\newcommand{\griz}{$g\,'r\,'i\,'z\,'\!$}
\newcommand{\JHK}{$JHK_{\rm s}$}
\def\ra#1#2#3{#1$^{\rm h}$#2$^{\rm m}$#3$^{\rm s}$}
\def\dec#1#2#3{$#1^\circ#2'#3''$}

\begin{document} 
  
\title{Illuminating the Darkest Gamma-Ray Bursts with Radio
Observations}

\author{
B.~A.~Zauderer\altaffilmark{1},
E.~Berger\altaffilmark{1},
R.~Margutti\altaffilmark{1},
A.~J.~Levan\altaffilmark{2}, 
F.~Olivares\altaffilmark{3},
D.~A.~Perley\altaffilmark{4,}\altaffilmark{5},
W.~Fong\altaffilmark{1},
A.~Horesh\altaffilmark{4},
A.~C.~Updike\altaffilmark{6},  
J.~Greiner\altaffilmark{3},
N.~R.~Tanvir\altaffilmark{7},
T.~Laskar\altaffilmark{1},
R.~Chornock\altaffilmark{1}, 
A.~M.~Soderberg\altaffilmark{1},
K.~M.~Menten\altaffilmark{8},
E.~Nakar\altaffilmark{9},
J.~Carpenter\altaffilmark{4},
and P.~Chandra\altaffilmark{10}
}
  
\altaffiltext{1}{Department of Astronomy, Harvard University,
Cambridge, MA 02138}

\altaffiltext{2}{Department of Physics, University of Warwick,
Coventry CV4 7AL, UK}

\altaffiltext{3}{Max-Planck-Institut f\"{u}r extraterrestrische
Physik, Giessenbachstra$\beta$e, 85748 Garching, Germany}

\altaffiltext{4}{Division of Physics, Mathematics and Astronomy,
California Institute of Technology, Pasadena, CA 91225}

\altaffiltext{5}{Hubble Fellow}

\altaffiltext{6}{Department of Chemistry and Physics, 
Roger Williams University, Bristol, RI 02809}

\altaffiltext{7}{Department of Physics and Astronomy, University 
of Leicester, Leicester, LE1 7RH, UK}

\altaffiltext{8}{Max-Planck-Institut f\"{u}r Radioastronomie,
53121 Bonn, Germany}

\altaffiltext{9}{Department of Astrophysics, Sackler School of Physics
and Astronomy, Tel Aviv University, 69978 Tel Aviv, Israel}

\altaffiltext{10}{Division of Physics, Royal Military College of Canada,
Kingston, Ontario K7K7B4, Canada}

\begin{abstract} We present X-ray, optical, near-infrared (IR), and radio
observations of GRBs 110709B and 111215A, as well as optical and
near-IR observations of their host galaxies.  The combination of X-ray
detections and deep optical/near-IR limits establish both bursts as
``dark''.  Sub-arcsecond positions enabled by radio detections lead to
robust host galaxy associations, with optical detections that indicate
$z\lesssim 4$ (110709B) and $z\approx 1.8-2.7$ (111215A).  We
therefore conclude that both bursts are dark due to substantial
rest-frame extinction.  Using the radio and X-ray data for each burst
we find that GRB\,110709B requires $A_V^{\rm host}\gtrsim 5.3$ mag and
GRB\,111215A requires $A_V^{\rm host}\gtrsim 8.5$ mag ($z=2$).  These
are among the largest extinction values inferred for dark bursts to
date.  The two bursts also exhibit large neutral hydrogen column
densities of $N_{\rm H,int} \gtrsim 10^{22}$ cm$^{-2}$ ($z=2$) as
inferred from their X-ray spectra, in agreement with the trend for
dark GRBs.  Moreover, the inferred values are in agreement with the
Galactic $A_V-N_H$ relation, unlike the bulk of the GRB population.
Finally, we find that for both bursts the afterglow emission is best
explained by a collimated outflow with a total beaming-corrected
energy of $E_\gamma+E_K\approx (7-9)\times 10^{51}$ erg ($z=2$)
expanding into a wind medium with a high density, $\dot{M}\approx
(6-20)\times 10^{-5}$ M$_\odot$ yr$^{-1}$ ($n\approx 100-350$
cm$^{-3}$ at $\approx 10^{17}$ cm).  While the energy release is
typical of long GRBs, the inferred density may be indicative of larger
mass loss rates for GRB progenitors in dusty (and hence metal rich)
environments.  This study establishes the critical role of radio
observations in demonstrating the origin and properties of dark GRBs.
Observations with the JVLA and ALMA will provide a sample with
sub-arcsecond positions and robust host associations that will help to
shed light on obscured star formation and the role of metallicity in
GRB progenitors.  \end{abstract}


\section{Introduction}
\label{sec:intro}

Long-duration gamma-ray bursts (GRBs) have been linked to the deaths of
massive stars, and hence to star formation activity, through their
association with star forming galaxies (e.g.,
\citealt{dkb+98,fls+06,wbp+07}) and with Type Ic supernova explosions
(e.g., \citealt{WoosleyBloom06}).  Across a wide range of cosmic
history a substantial fraction of the star formation activity ($\sim
70\%$ at the peak of the star formation history, $z\sim 2-4$) is
obscured by dust, with about 15\% of the total star formation rate
density occurring in ultra-luminous infrared galaxies (e.g.,
\citealt{bif+09,rs09,mcd+11}).  As a result, we expect some GRBs to
occur in dusty environments that will diminish or completely
extinguish their optical (and perhaps even near-IR) afterglow
emission.  Such events can be used as signposts for the locations and
relative fraction of obscured star formation across a wide redshift
range.  In addition, they can provide insight into the role of
metallicity in GRB progenitors since dusty environments generally
require substantial metallicity.

These so-called optically-dark GRBs are indeed known to exist (e.g.,
\citealt{ggp+98,dfk+01,fjg+01,pfg+02}), but the lack of an optical detection
does not necessarily point to dust obscuration.  Most prosaically, the
lack of detected optical emission may be due to inefficient follow-up
observations, or to intrinsically dim events (e.g., \citealt{bkb+02}).
Another potential origin of dark bursts is a high redshift, with the
optical emission suppressed by Ly$\alpha$ absorption at $\lambda_{\rm
obs}\lesssim 1216\,$\AA\,$\times (1+z)$
\citep{hnr+06,sdc+09,tfl+09,clf+11}.  Such events are clearly of great
interest since a dropout above $\sim 1$ $\mu$m points to $z\gtrsim 7$,
while events which are also dark in the near-IR can potentially arise
at $z\gtrsim 18$ ($K$-band dropout).  Naturally, high redshift bursts
will also lack host galaxy detections in the optical band.  On the
other hand, their afterglow emission redward of the Ly$\alpha$ break
will follow the expected synchrotron spectrum
($F_\nu\propto\nu^{-\beta}$) with $\beta\approx 0.5-1$ \citep{spn98}.

To determine whether a burst is genuinely dark, due to extinction or
to a high redshift rather than to an inefficient search, it is common
to compare the observed limits with the expected optical/near-IR
brightness based on the X-ray afterglow.  This approach relies on the
simple power law shape of the afterglow synchrotron emission.  One
definition of dark bursts uses an optical to X-ray spectral index of
$\beta_{\rm OX} \lesssim 0.5$ ($\beta$ is the slope of the synchrotron 
spectrum in log space; see \citealt{jakobsson+04} for details), since this is the
shallowest expected slope in the standard afterglow model.  A variant of this
condition uses knowledge of the X-ray spectral index ($\beta_{\rm
X}$), and defines bursts as dark if $\beta_{\rm OX}-\beta_{\rm
X}\lesssim -0.5$ \citep{hkg+09}, since this is the shallowest expected
relative slope.  These definitions can reveal evidence for dust
extinction even if an optical afterglow is detected.

This approach has been used by several groups to identify and study
dark bursts.  In the pre-\swift\ sample \citet{jakobsson+04} found
that about 10\% of the bursts are dark, while \citet{kkz06} found a
mean extinction of $A_V^{\rm host}\approx 0.2$ mag, indicating that
most events are not dark.  \citet{Schady+07} studied several bursts
with \swift\ X-ray and UV/optical detections and found a mean
extinction level of $A_V^{\rm host}\approx 0.3$ mag, and that bursts
which are dark blueward of $V$-band are likely to have rest-frame
extinction about an order of magnitude larger.  \citet{mmk+08} used
rapid optical observations of 63 \swift\ bursts and found that about
50\% showed evidence of mild extinction.  A similar conclusion was
reached by \citet{Cenko+09} based on rapid optical observations of 29
\swift\ bursts, with an 80\% detection fraction, but with roughly half
exhibiting suppression with respect to the X-ray emission.  A
follow-up study of the latter sample by \citet{perley+09} aimed at
identifying host galaxies in the optical (thereby ruling out a high
redshift origin) and found that $\lesssim 7\%$ of \swift\ bursts are
located at $z\gtrsim 7$.  The majority of the dark bursts in their
sample instead require $A_V^{\rm host}\gtrsim 1$ mag, with a few cases
reaching $\sim 2-6$ mag.  \citet{Melandri+12} studied a complete
sample of 58 bright \swift\ bursts, of which 52 have known redshifts
and found that the fraction of dark bursts is about 30\%, mainly due
to extinction (nearly all the dark bursts in their sample have
$z\lesssim 4$).  Thus, the afterglow emission of $\sim 1/3-1/2$ of all GRBs are affected by
dust, although in most cases the required extinction is modest,
$A_V^{\rm host}\approx 0.3$ mag \citep{kkz06,Schady+07,pcb+09,gkk+11}.
There are only a few known cases with large extinction of $A_V^{\rm
host}\sim {\rm few}$ mag.

Concurrent studies of the neutral hydrogen column density
distribution, inferred from the X-ray afterglow spectra, suggest that
dark bursts exhibit systematically larger values of $N_{\rm H,int}
\gtrsim 10^{22}$ cm$^{-2}$ compared to bursts with little or no
extinction, which have a median of $N_{\rm H,int}\approx 4\times
10^{21}$ cm$^{-2}$ \citep{csm+12,Margutti+12}.  On the other hand, the
measured extinction for GRBs is generally lower than expected based on
the values of $N_{\rm H,int}$ and the Galactic $A_V-N_H$ relation,
previously attributed to dust destruction by the bright X-ray/UV
emission (e.g., \citealt{gw01,Schady+07,pcb+09}).

Finally, the host galaxies of at least some dark bursts appear to be
redder, more luminous, more massive, and more metal rich than the
hosts of optically-bright GRBs
\citep{bfk+07,pcb+09,lkg+10,pmu+11,kgs+11}, potentially indicating
that the extinction is interstellar in origin rather than directly
associated with the burst environment.
One specific example is GRB\,030115, not only one of
the first examples of a heavily extinguished GRB afterglow, but residing
in a host classified as an extremely red object (ERO; \citealt{lfr+06}).
Some host galaxies of optically-bright bursts have been detected in the radio and
(sub)millimeter ranges (e.g., \citealt{tbb+04}, indicating obscured star 
formation rates of $\sim 10^2-10^3$ M$_\odot$ yr$^{-1}$ \citep{bck+03}.  
This suggests that the dust distribution within the host galaxies is patchy 
and that not all GRBs in dusty host galaxies are necessarily dust-obscured
\citep{bck+03,pcb+09}.  
Further work by \citet{slt+12} indicates that dark GRB host galaxies may
be systematically redder and more massive than optically bright GRB
host galaxies and have lower metallicity than sub-millimeter galaxies, 
leading to prior biases in GRB host galaxy population studies.

Here we present multi-wavelength observations that reveal two of the
darkest known bursts to date, GRBs 110709B and 111215A.  These events 
have such large rest-frame extinction that they lack any afterglow
detection in the optical and near-IR bands (to $\lambda_{\rm
obs}\approx 2.2$ $\mu$m) despite rapid follow-up.  However, both
events are detected in the radio, providing sub-arcsecond positions
and secure associations with host galaxies detected in the optical.
This allows us to rule out a high redshift origin, while the
combination of radio and X-ray data place robust lower limits on the
rest-frame extinction, with $A_V^{\rm host}\gtrsim 5.3$ and $\gtrsim
8.5$ mag (for $z=2$), respectively.  The radio and X-ray data also
allow us to determine the explosion properties and the circumburst
densities.  This study demonstrates that radio detections with the JVLA,
and soon the Atacama Large Millimeter/submillimeter Array (ALMA), 
provide a promising path to accurate localization of
dark GRBs, robust estimates of the extinction in the absence of any
optical/near-IR detections, and a comparative study of their explosion
properties and parsec-scale environments.

The plan of the paper is as follows.  We describe the multi-wavelength
observations of GRBs 110709B and 111215A and their host galaxies in
\S\ref{sec:obs}.  These observations establish that both events are
dark due to extinction.  We model the radio and X-ray data to extract
the minimum required extinction, as well as the explosion and
circumburst properties in \S\ref{sec:model}.  In \S\ref{sec:disc} we
place the large inferred extinction values and the inferred burst
properties within the framework of existing samples of
optically-bright and dark bursts.  We summarize the results and note
key future directions in \S\ref{sec:conc}.  Throughout the paper we
report magnitudes in the AB system (unless otherwise noted) and use
Galactic extinction values of $E(B-V)\approx 0.044$ mag for
GRB\,110709B and $E(B-V)\approx 0.057$ mag for GRB\,111215A
\citep{sf11}.  We use the standard cosmological parameters: $H_{0}=71$
km s$^{-1}$ Mpc$^{-1}$, $\Omega_{\Lambda}= 0.73$, and $\Omega_{\rm
M}=0.27$.

\section{Observations}
\label{sec:obs}

\subsection{Discovery and Burst Properties}

\subsubsection{GRB\,110709B}

GRB\,110709B was discovered on 2011 July 9 at 21:32:39 UT
\citep{gcn12122} by the \swift\ Burst Alert Telescope (BAT; 15$-$150 keV;
\citealt{Bart+05}), and by Konus-WIND (20 keV$-$5 MeV; \citealt{Golentskii+11}). 
\swift\ X-ray Telescope (XRT; 0.3$-$10 keV; \citealt{Burrows05}) observations 
commenced at $\delta t\approx 80$ s \citep{gcn12122}, localizing the X-ray
afterglow to RA=\ra{10}{58}{37.08}, Dec=\dec{-23}{27}{17.6} (J2000),
with an uncertainty of $1.4''$ (90\% containment, UVOT-enhanced;
\citealt{gcn12136}).  \swift\ UV/Optical Telescope (UVOT;
\citealt{Roming05}) observations began at $\delta t\approx 90$ s, but
no afterglow candidate was identified to a $3\sigma$ limit of $\approx
21.1$ mag in the white filter \citep{gcn12122}.  A large flare was
detected with the BAT, XRT, and Konus-WIND at $\delta t\approx 11$ min
with an intensity comparable to the main event \citep{gcn12124}.

\citet{zhang+11} presents a detailed analysis of the prompt emission.  
The initial trigger had a duration and fluence in
the BAT $15-150$ keV band of $T_{90}\approx 56$ s and
$F_\gamma=(8.95^{+0.29}_{-0.62})\times 10^{-6}$ erg cm$^{-2}$, while
the flare had $T_{90}\approx 259$ s and
$F_\gamma=(1.34^{+0.05}_{-0.07})\times 10^{-5}$ erg cm$^{-2}$
\citep{zhang+11}.  No emission was detected in the BAT band at $\delta
t\approx 180-485$ s.  Combining both events, the resulting duration is
$T_{90}\approx 846\pm6$ s \citep{gcn12144}.  The initial trigger had a 
fluence in the Konus-WIND $20-5000$ keV band of 
$(2.6\pm 0.2)\times 10^{-5}$ erg cm$^{-2}$.  \citet{zhang+11} calculate
the total fluence in the \swift/XRT band to be 4.07 $\pm$ 0.56 $\times
10^{-6}$ erg cm$^{-2}$.

\subsubsection{GRB\,111215A}

GRB\,111215A was discovered on 2011 December 15 at 14:04:08 UT by the
\swift/BAT \citep{gcn12681}.  \swift/XRT observations began at $\delta
t\approx 409$ s and localized the X-ray afterglow to
RA=\ra{23}{18}{13.29}, Dec=\dec{+32}{29}{38.4}, with an uncertainty of
$1.4''$ (90\% containment, UVOT-enhanced; \citealt{gcn12690}).  No
afterglow candidate was identified by the UVOT to a $3\sigma$ limit of
$\gtrsim 21.7$ mag in the white filter \citep{gcn12693}.  The duration
and fluence are $T_{90}\approx 796$ s and $F_\gamma=(4.5\pm 0.5)\times
10^{-6}$ erg cm$^{-2}$ ($15-150$ keV; \citealt{gcn12689}).

\subsection{X-ray Observations}
\label{sec:xrays}

\subsubsection{GRB\,110709B}

We analyzed the XRT data using the HEASOFT package (v6.11) and
corresponding calibration files.  We utilized standard filtering and
screening criteria, and generated a count-rate light curve following
the prescriptions by \citet{Margutti10}.  The data were re-binned with
the requirement of a minimum signal-to-noise ratio of 4 in each
temporal bin.

\begin{figure}
\centering
\includegraphics[totalheight=0.8\textwidth]{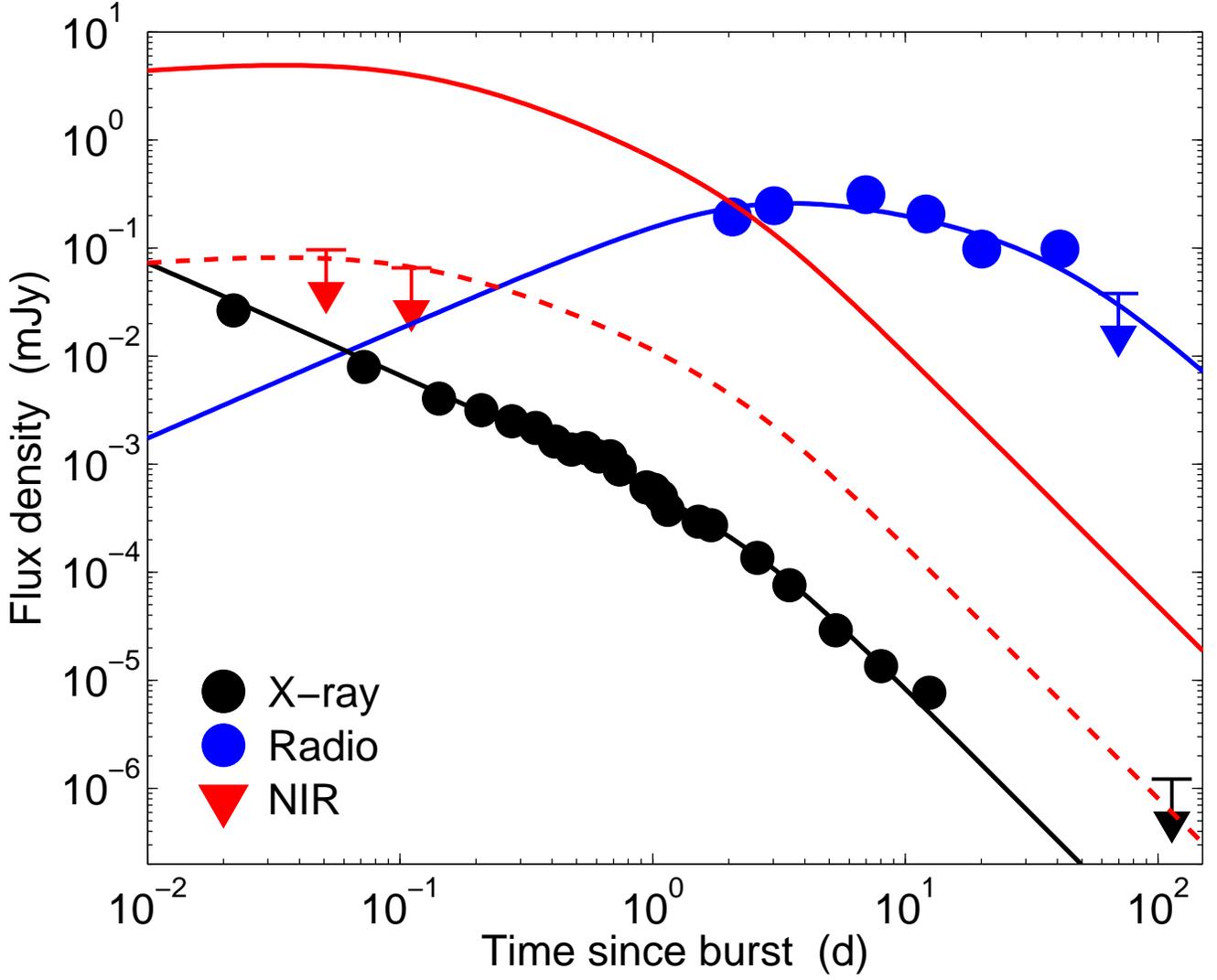}
\caption{X-ray (black) and radio (blue) light curves of GRB\,110709B,
with near-IR limits (red triangles).  For the purpose of display, the
XRT data are binned by orbit at $\delta t\lesssim 2$ d and by multiple
orbits thereafter.  The \chandra\ $3\sigma$ upper limit at $\delta
t\approx 113$ d is indicated with a black arrow.  Afterglow model fits
in each band are shown with solid lines.  To satisfy the near-IR upper
limits we require $A_V\gtrsim 3.4-10.5$ mag (for $z=4$ to $1$; dashed
red line).  The derived afterglow parameters for GRB\,110709B are
listed in Table~\ref{tab:110709model}.
\label{fig:110709lc}}
\end{figure}

The XRT data exhibit significant spectral evolution during the early
bright phase ($\delta t\lesssim 1$ ks), with spectral hardening when
the burst re-brightened at $\sim 11$ min.  Due to this variation, we
performed a time-resolved spectral analysis, accumulating signal over
time intervals defined to contain a minimum of about $1000$ photons.
Spectral fitting was done with {\tt Xspec} (v12.6; \citealt{arn96}), assuming a
photoelectrically absorbed power law model and a Galactic neutral
hydrogen column density of $N_{\rm H,MW}=5.6\times 10^{20}$ cm$^{-2}$
\citep{Kalberla05}.  We extracted a spectrum in a time interval when
no spectral evolution was apparent ($15-150$ ks) and used this to
estimate the contribution of additional absorption.  We find that the
data are best modeled by an absorbed power-law\footnote{$tbabs\times
ztbabs\times pow$} model with a spectral photon index of
$\Gamma=2.3\pm 0.1$ and excess absorption of $N_{\rm H,int}=(1.4\pm
0.2)\times 10^{21}$ cm$^{-2}$ at $z=0$ (90\% c.l., C-stat = 559 for
604 degrees of freedom).  The excess absorption was used as a fixed
parameter in the time-resolved spectral analysis, allowing us to
derive a count-to-flux conversion factor for each spectrum.  The
resulting unabsorbed $0.3-10$ keV flux light curve properly accounts
for spectral evolution of the source.  We note that for the purpose of
display in Figure~\ref{fig:110709lc} we binned the XRT data by orbit
at $\delta t\lesssim 2$ d and by multiple orbits thereafter.

We also observed GRB\,110709B for 15 ks with the Advanced CCD Imaging
Spectrometer (ACIS-S) on-board the {\it Chandra} X-ray Observatory on
2011 July 23.60 UT ($\delta t\approx 13.7$ d).  The afterglow is
detected with a count rate of $(4.1\pm 0.5)\times 10^{-3}$ s$^{-1}$
($0.5-8$ keV), corresponding to an unabsorbed $0.3-10$ keV flux of
$(6.0\pm 0.7)\times 10^{-14}$ erg s$^{-1}$ cm$^{-2}$ using the XRT
spectral parameters.  Relative astrometry with the {\it Chandra} position
will be discussed in \S 2.6. A second, 9 ks \emph{Chandra}/ACIS-S observation
was performed on 2011 October 31.83 UT ($\delta t\approx 113.4$ d).
The X-ray afterglow is not detected to a $3\sigma$ limit of $\lesssim
9\times 10^{-15}$ erg s$^{-1}$ cm$^{-2}$, well above the extrapolation
of the early light curve to this epoch (Figure~\ref{fig:110709lc}).

At $\delta t\gtrsim 12$ d the XRT light curve flattens to a level of
about $10^{-13}$ erg s$^{-1}$ cm$^{-2}$.  The {\it Chandra}
observations reveal that this is due to a contaminating source located
about $3.8''$ from the GRB position, well within the point spread
function (PSF) of XRT.  We subtract the flux of this source, $(6.0\pm
0.7)\times 10^{-14}$ erg s$^{-1}$ cm$^{-2}$, from the XRT light curve
of GRB\,110709B in Figure~\ref{fig:110709lc}.\footnote{We note that 
the flux from the contaminating source is constant between the two 
epochs of {\em Chandra}, and rules out any origin as a lensed 
counterpart of the GRB, which may have caused the two $\gamma$-ray 
triggers.}

\subsubsection{GRB\,111215A}

\begin{figure}
\centering
\includegraphics[totalheight=0.8\textwidth]{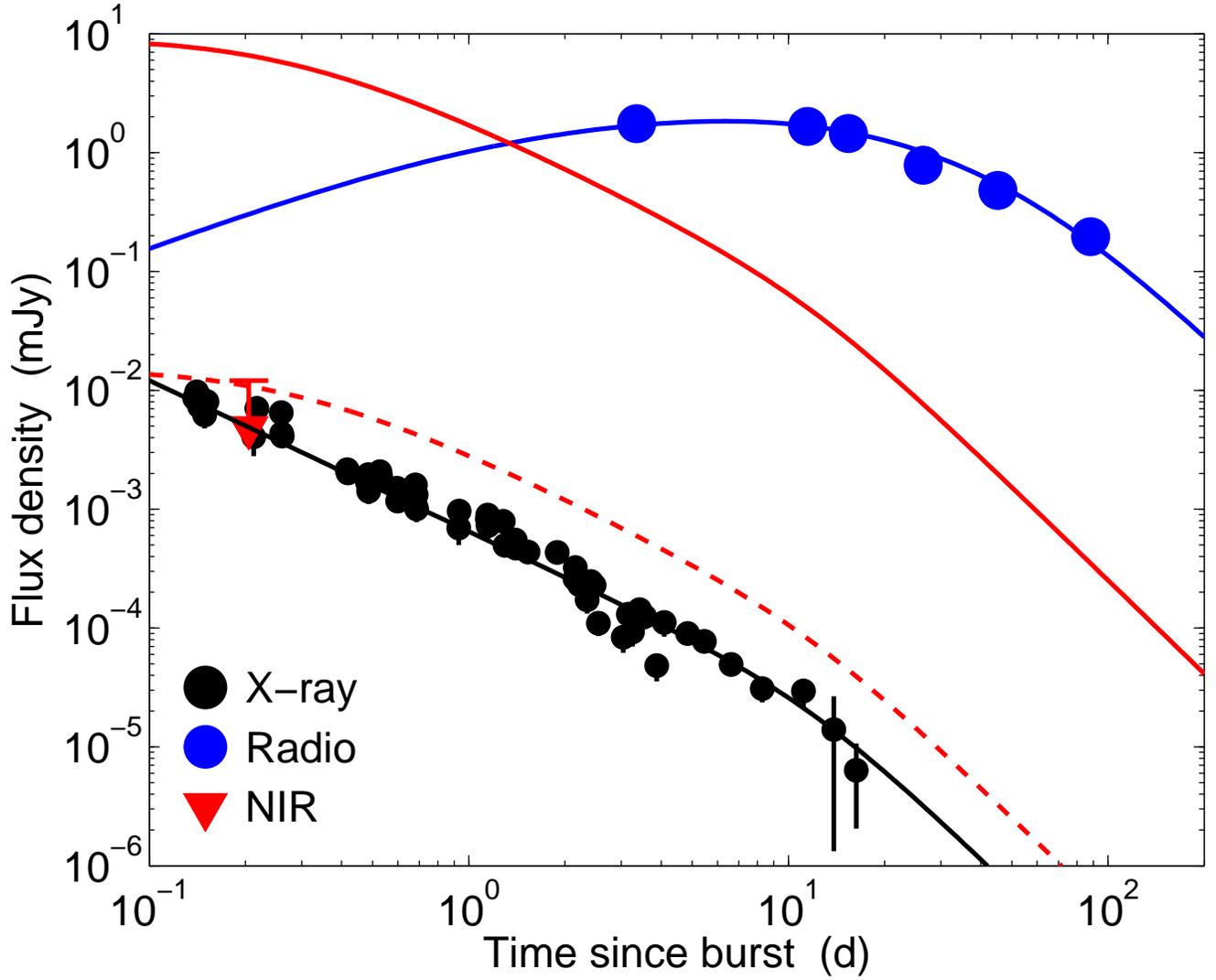}
\caption{X-ray (black) and radio (blue) light curves of GRB\,111215A,
with near-IR limits (red triangle).  Afterglow model fits in each band
are shown with solid lines.  To explain the near-IR upper limit we
require $A_V\gtrsim 8.5$ mag (dashed line, assuming $z=2$).  The
derived afterglow parameters are listed in
Table~\ref{tab:111215model}.
\label{fig:111215lc}}
\end{figure}

We analyzed the XRT data for GRB\,111215A in the same manner described
above.  At $\delta t\lesssim 2$ ks the light curve exhibits flares
with a hard-to-soft spectral evolution.  To obtain a reliable estimate
of the intrinsic neutral hydrogen absorption in addition to the
Galactic value ($N_{\rm H,MW}=5.5\times 10^{20}$ cm$^{-2}$;
\citealt{Kalberla05}), we extracted a spectrum collecting the photon
count data in the time interval $5-2000$ ks, when no spectral
evolution is apparent.  The data are best modeled by an absorbed
power-law with $\Gamma=2.2\pm 0.1$ and $N_{\rm H,int}=(3.1\pm
0.4)\times 10^{21}$ cm$^{-2}$ at $z=0$ (90\% c.l., C-stat = 499.13 for
512 degrees of freedom).  The resulting unabsorbed $0.3-10$ keV flux
light curve is shown in Figure~\ref{fig:111215lc}.

\subsection{Optical/Near-IR Afterglow Limits}
\label{sec:opt}

\subsubsection{GRB\,110709B}

We observed GRB\,110709B with the Gemini Multi-Object Spectrograph
(GMOS; \citealt{Hook+2004}) on the Gemini-South 8-m telescope in
$r$-band on 2011 July 10.03 and 13.96 UT ($\delta t\approx 3.2$ hr and
$\approx 4.1$ d).  We analyzed the data with the {\tt gemini} package
in IRAF, but did not detect any sources within the XRT error circle to
$3\sigma$ limits of $\gtrsim 23.8$ mag ($\delta t\approx 3.2$ hr) and
$\gtrsim 25.1$ mag ($\delta t\approx 4.1$ d).  A comparison of the
early $r$-band limit ($F_{\rm \nu,opt} \lesssim 1.1$ $\mu$Jy) to the
X-ray flux density at the same time ($F_{\rm \nu,X} \approx 4.0$
$\mu$Jy) indicates an optical to X-ray spectral index $\beta_{\rm
OX}\lesssim -0.2$.  This is substantially flatter than the minimum
expected spectral index of $\beta_{\rm OX}=0.5$, indicating that
GRB\,110709B is a dark burst; for $\beta_{\rm OX}=0.5$ the expected
$r$-band brightness is $\approx 19$ mag.  Similarly, relative to the
X-ray spectral index we find $\beta_{\rm OX}-\beta_{\rm X} \lesssim
-1.5$, clearly satisfying the dark burst condition $\beta_{\rm
OX}-\beta_{\rm X}\lesssim -0.5$.

We also observed the burst with the Gamma-Ray Burst
Optical/Near-Infrared Detector (GROND; \citealt{Greiner07,Greiner08})
mounted on the Max Planck Gesellschaft / European Southern Observatory
2.2-m telescope at La Silla Observatory starting on 2011 July 9.94 UT
($\delta t\approx 1$ hr) simultaneously in \griz\JHK, with an average
seeing of $1.5''$ and an average airmass of 1.3.  We analyzed the data
using standard {\tt pyraf/IRAF} tasks, following the procedures
described in \citet{Kruehler08}.  We calibrated the optical channels
with zero-points computed during photometric conditions, and the
near-IR channels using stars in the 2MASS catalog.  Calibration
uncertainties vary in the range $0.02-0.12$ mag for \JHK, and we add
these systematic errors with the statistical errors in quadrature.  We
do not detect any source in the individual or stacked images within
the XRT error circle.  The upper limits are listed in
Table~\ref{tab:grond}, derived by forcing the photometry at the
position of the radio counterpart (\S\ref{sec:evla}).  A comparison of
the $K_{\rm s}$ band limit at $\delta t\approx 2.7$ hr ($F_{\rm
\nu,NIR}\lesssim 69$ $\mu$Jy) to the X-ray flux density at the same
time indicates a near-IR to X-ray slope of $\lesssim 0.35$, indicating
that GRB\,110709B is a dark burst in the near-IR as well.

The absence of optical/near-IR emission can be due to dust extinction,
or alternatively to a high redshift, $z\gtrsim 18$ based on the
$K_{\rm s}$-band non-detection.  We rule out a high redshift
origin in \S \ref{sec:host}.

\subsubsection{GRB\,111215A}

Several optical afterglow searches on timescales of $\delta t\approx
7$ min to $6.8$ hr led to non-detections of an afterglow with limits
of $\gtrsim 17.5-22.8$ mag in various filters
\citep{gcn12682,gcn12683,gcn12685,gcn12686,gcn12687,gcn12688,gcn12703}.
The deepest optical limit is $m_R\gtrsim 22.8$ mag at a mid-time of
$\delta t\approx 37$ min \citep{gcn12683}, corresponding to a flux
density of $F_{\rm \nu,opt}\lesssim 2.3$ $\mu$Jy.  Near-IR
observations also led to non-detections \citep{gcn12695,gcn12696},
with the deepest limit being $m_K\gtrsim 19.3$ mag (Vega) at $\delta
t\approx 0.20$ d \citep{gcn12696}, corresponding to $F_{\rm \nu,NIR}
\lesssim 12$ $\mu$Jy.

A comparison of the deepest optical limit to the X-ray flux density at
the same time ($F_{\rm \nu,X}\approx 54$ $\mu$Jy) indicates
$\beta_{\rm OX}\lesssim -0.5$.  Similarly, a comparison of the deepest
near-IR limit to the X-ray flux density at the same time ($F_{\rm
\nu,X} \approx 4.2$ $\mu$Jy) indicates $\beta_{\rm NIRX}\lesssim
0.15$.  In comparison to the X-ray spectral index we find $\beta_{\rm
OX}-\beta_{\rm X}\lesssim -1.7$ and $\beta_{\rm NIRX}-\beta_{\rm
X}\lesssim -1.05$.  Thus, GRB\,111215A is a dark burst in the optical
and near-IR.  As in the case of GRB\,110709B, the absence of
optical/near-IR emission can be due to dust extinction, or
alternatively to a high redshift, $z\gtrsim 18$ based on the $K_{\rm
s}$-band non-detection.

\subsection{JVLA Centimeter Observations}
\label{sec:evla} 

\subsubsection{GRB\,110709B}

We observed GRB\,110709B with the NRAO Karl G.~Jansky Array (JVLA) 
beginning on 2011 July 11.97 UT
($\delta t\approx 2.1$ d) at a mean frequency of 5.8 GHz and detected
a single, unresolved radio source within the XRT error circle.  This
detection provided the first accurate position for the burst.
Follow-up observations demonstrated that the source initially
brightened and subsequently faded away, establishing it as the radio
afterglow of GRB\,110709B.

All observations utilized the WIDAR correlator \citep{rperley+11} with
$\sim 2$ GHz bandwidth.  We calibrated and analyzed the data using
standard procedures in the Astronomical Image Processing System (AIPS;
\citealt{greisen03}).  We excised edge channels and channels affected
by radio frequency interference, reducing the effective bandwidth by
$\sim 25\%$ at 5.8 GHz.  Due to the low declination of the source we
also excised data when $>2$ m of a given antenna was shadowed by
another antenna.  For the observation at 21.8 GHz we performed
reference pointing at 8.4 GHz and applied the pointing solutions, per
standard high frequency observing procedures.  We observed 3C286 for
band-pass and flux calibration and interleaved observations of
J1048$-$1909 for gain calibration every 3~m at 21.8 GHz and 
J1112$-$2158 every 4~m at 5.8 GHz.  The resulting flux densities at 
5.8 and 21.8 GHz are listed in Table~\ref{tab:evla1}.  The uncertainties 
are $1\sigma$ statistical errors, and we note an additional uncertainty in 
the absolute flux scaling of $\sim 5\%$.  The 5.8 GHz light curve is shown in
Figure~\ref{fig:110709lc}.

Finally, to determine the position of the radio afterglow we fit the
source in each image with a Gaussian profile (AIPS task JMFIT), and
calculated the mean radio position weighted by the resulting
statistical uncertainties.  We find a weighted mean position of
RA=\ra{10}{58}{37.113} ($\pm 0.001$), Dec=\dec{-23}{27}{16.76} ($\pm
0.02$).

\subsubsection{GRB\,111215A}

\begin{figure}
\centering
\includegraphics[totalheight=0.8\textwidth]{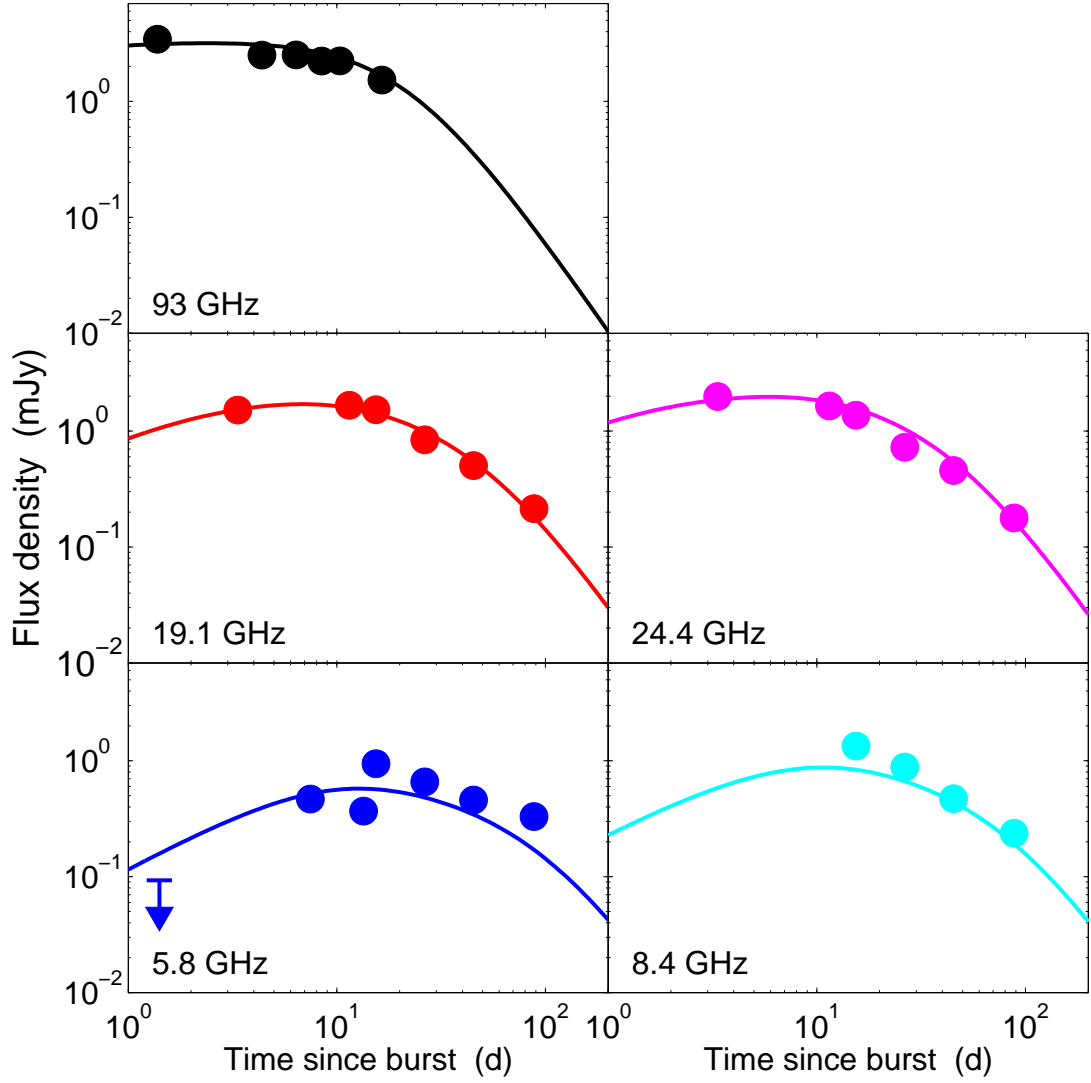}
\caption{Radio light curves of GRB\,111215A at 5.8, 8.4, 19.1, 24.4,
and 93 GHz.  Afterglow model fits are shown with the solid lines.  The
derived afterglow parameters are listed in
Table~\ref{tab:111215model}.
\label{fig:111215lc2}} 
\end{figure}

We observed GRB\,111215A with the JVLA beginning on 2011 December
17.00 UT ($\delta t\approx 1.4$ d), and subsequently detected the
radio afterglow starting at $\delta t\approx 3.35$ d.  Observations
between 1.4 and 140 days were obtained at mean frequencies of 5.8, 8.4
and 21.8 GHz.  We used 3C48 for band-pass and flux calibration and
interleaved observations of J2311$+$3425 every $\sim$4~m for gain calibration 
at 21.8 GHz and J2340$+$2641 every $\sim$5.5~m for gain calibration at 5.8 
and 8.4 GHz.  The
data were analyzed in the same manner described above, and the
resulting flux densities are listed in Table~\ref{tab:evla2}.
The light curves are shown in Figures~\ref{fig:111215lc} and
\ref{fig:111215lc2}.

We obtain a weighted mean position from all 21.8 GHz detections of
RA=\ra{23}{18}{13.314} ($\pm 0.004$), Dec=\dec{+32}{29}{39.07} ($\pm
0.06$).  The JVLA data (along with CARMA observations: \S\ref{sec:mm})
provide the only sub-arcsecond position for GRB\,111215A.

\subsection{Millimeter Observations}
\label{sec:mm}

\subsubsection{GRB\,110709B}

GRB\,110709B was observed at a frequency of $345$ GHz with the Large
APEX Bolometer Camera (LABOCA; \citealt{LABOCA}) on the Atacama
Pathfinder Experiment (APEX) telescope beginning on 2011 July 11.90 UT
($\delta t\approx 2.0$ d).  No submillimeter counterpart was detected
to a $3\sigma$ upper limit of $\lesssim 6.9$ mJy \citep{gcn12151}.

\subsubsection{GRB\,111215A}

We observed GRB\,111215A with the Submillimeter Array (SMA;
\citealt{Ho+04}) beginning on 2011 December 18.15 UT ($\delta t\approx
2.7$ d) at 230 GHz with 8 GHz bandwidth.  We performed gain
calibration using J2311$+$344 and J2236$+$284, band-pass calibration
using 3C84, and flux calibration using real-time measurements of the
system temperatures, with observations of Uranus and Callisto to set
the overall flux scale (accurate to about $10\%$).  We calibrated the
data using the MIR software package, and analyzed and imaged the
observation with MIRIAD.  We do not detect the GRB afterglow to a
$3\sigma$ limit of $\lesssim 2.6$ mJy.

We also observed GRB\,111215A with the Combined Array for Research in
Millimeter Astronomy (CARMA) beginning on 2011 December 16.97 UT
($\delta t\approx 1.38$ d) at 93 GHz, and continued observations for
$\approx 16$ d; see Table~\ref{tab:evla2}.  We used Neptune as
the primary flux calibrator, and J1824$+$568 and J1638$+$573 as
band-pass and gain calibrators, respectively.  The overall uncertainty
in the absolute flux calibration is $\sim 15\%$.  We calibrated and
visualized the data with the MIRIAD software package
\citep{miriad+95,miriad+11}, and measured the flux densities with AIPS
for consistency with the JVLA observations
(Table~\ref{tab:evla2}).

We fit the position of the detected mm afterglow in each epoch with a
Gaussian fixed to the beam size, and compute the mean position,
weighted by the uncertainty in each epoch, leading to
RA=\ra{23}{18}{13.321} ($\pm 0.007$) and Dec=\dec{+32}{29}{39.04}
($\pm 0.07$).  This position is consistent with the \swift/X-ray position.

\subsection{Host Galaxy Observations and Redshift Constraints}
\label{sec:host}

\subsubsection{GRB\,110709B}

We obtained Hubble Space Telescope ({\it HST}) observations of
GRB\,110709B with the Wide Field Camera 3 (WFC3;
\citealt{Kimble+2008}) on 2011 November 8.94 UT ($\delta t\approx 122$
d) using the ultra-violet imaging spectrograph (UVIS) with the F606W
filter, and on 2011 November 12.94 UT ($\delta t\approx 126$ d) using
the infrared (IR) channel with the F160W filter (PI: Levan).  One
orbit of observations was obtained in each filter, with exposure times
of 2610 s (F606W) and 2480 s (F160W).  The data were processed using
{\tt multidrizzle} \citep{fh02,kfh+06} with output pixel scales set at
$0.02''$ and $0.07''$, respectively.

To locate the absolute radio and X-ray afterglow positions on the {\it
HST} images we tie the F606W and F160W images to the 2MASS reference
frame using a wider field Gemini-South GMOS $r$-band image as an
intermediary.  Using 20 common sources with 2MASS, the absolute
astrometry of the Gemini image has an rms scatter of $\sigma_{\rm
GMOS-2MASS}=0.15''$ in each coordinate.  The {\it HST} images are tied
to the GMOS image using 30 common sources with a resulting rms scatter
of $\sigma_{HST-{\rm GMOS}}=0.03''$ in each coordinate.  We further
refine the {\it Chandra} afterglow position relative to the {\it HST}
astrometric frame using a single common source.  This leads to a shift
in the X-ray position of $\delta{\rm RA}=-0.10''$ ($-0.07''$) and
$\delta{\rm Dec}=+0.20''$ ($+0.18''$) relative to the F606W (F160W)
reference frame.  There are no common sources between the {\it HST}
and JVLA images.

\begin{figure}
\centering
\includegraphics[totalheight=0.44\textwidth]{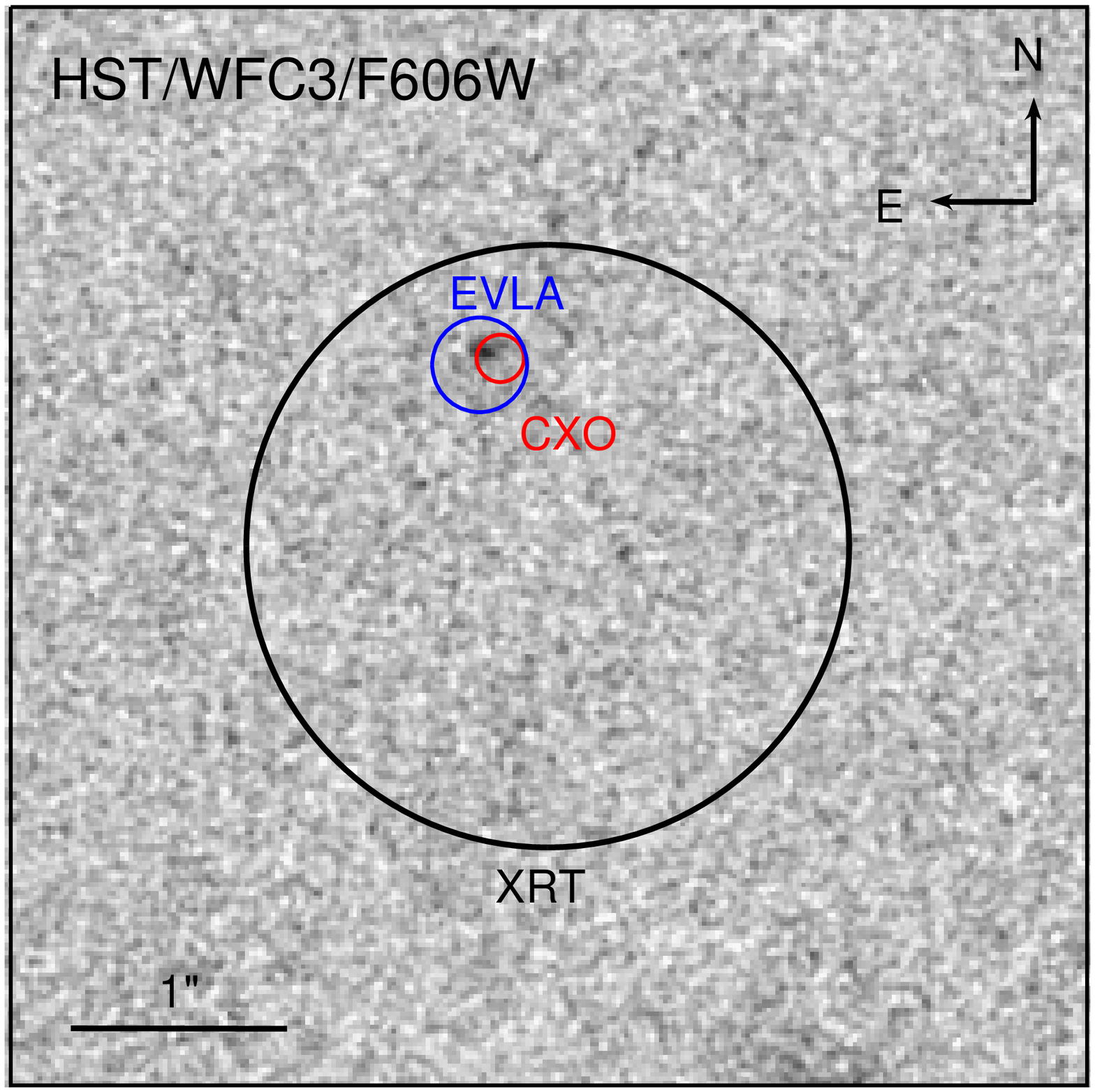}\hfill
\includegraphics[totalheight=0.44\textwidth]{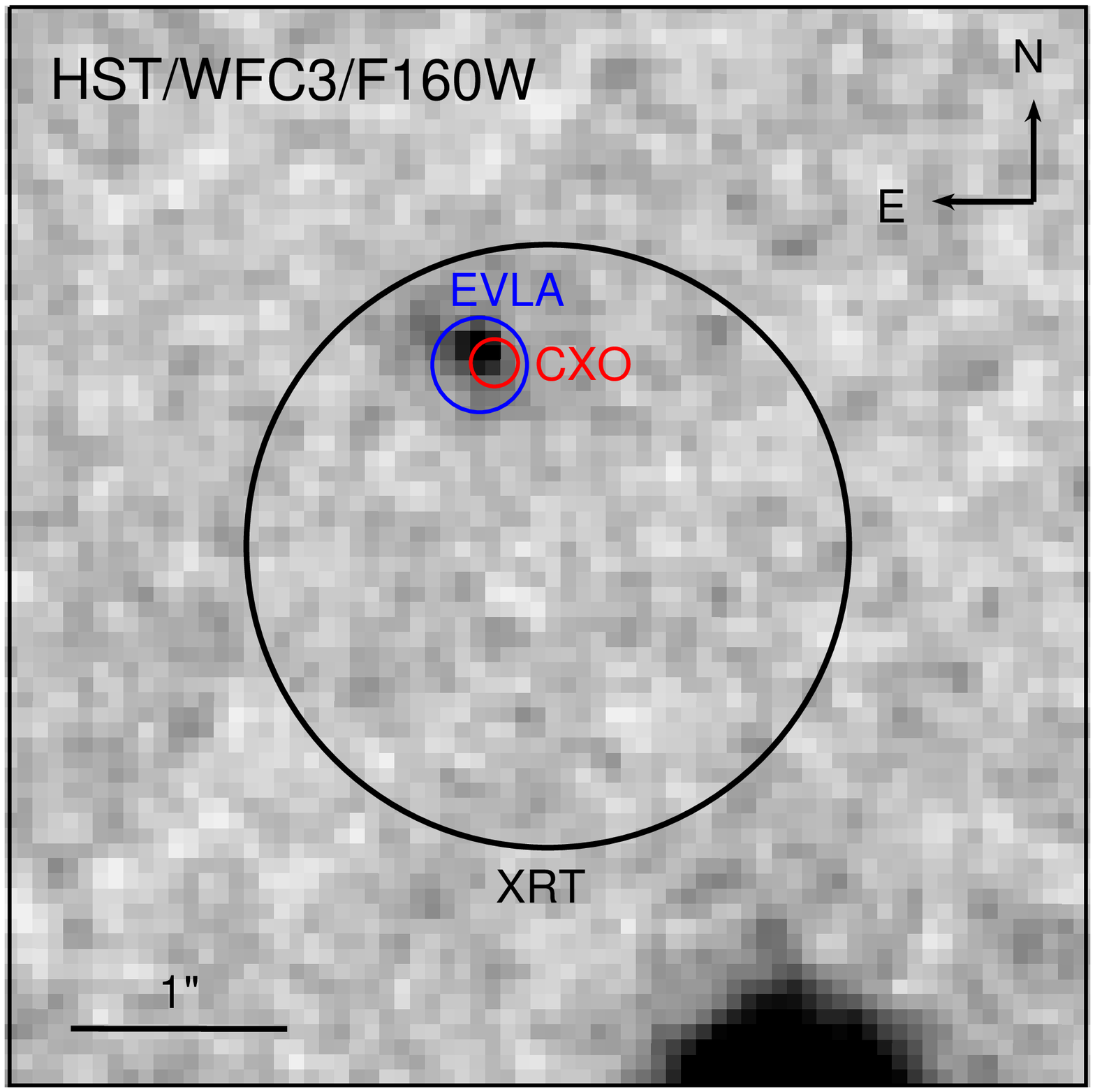}
\caption{{\it HST}/WFC3 images of the host galaxy of GRB\,110709B.
The XRT error circle ($1.4''$ radius; 90\% containment) is indicated
with the black circle.  The JVLA afterglow position is shown with a
blue circle ($0.22''$ radius; $1\sigma$ systematic), while the \chandra\ position
is shown with a red circle ($0.10''$; $1\sigma$).  The JVLA and
\chandra\ positions coincide with a galaxy that we consider to be the
host of GRB\,110709B. 
\label{fig:110709host}} 
\end{figure}

The resulting uncertainty in the radio afterglow position in the {\it
HST} reference frame is dominated by the absolute astrometry, and
corresponds to a radius of $0.22''$ ($1\sigma$; the centroid
uncertainty of the radio position is only $0.02''$ in each
coordinate).  The uncertainty in the X-ray afterglow position is
reduced by the relative tie of the \chandra\ and {\it HST} reference
frames, leading to a radius of $0.10''$ ($1\sigma$; the centroid
uncertainty of the \chandra\ position is about $0.08''$ in each
coordinate).  Within the uncertainty regions we find a single galaxy
in both {\it HST} images (Figure~\ref{fig:110709host}), which we
identify as the host of GRB\,110709B.  Photometry of this source using
the tabulated WFC3 zeropoints gives $m_{\rm F606W}=26.78\pm 0.17$ AB
mag and $m_{\rm F160W}=25.13\pm 0.10$ mag, corrected for Galactic
extinction.

To determine the probability of chance coincidence, $P_{\rm cc}
(<\delta R)=1-{\rm exp}[-\pi (\delta R)^2\sigma(\le m)]$, we use the
galaxy number counts, $\sigma(\le m)=10^{0.33(m-24)-2.32}$
arcsec$^{-2}$ \citep{hpm+97,bsk+06}.  For the XRT error circle we
infer $P_{cc}\approx 0.1$, while for the more precise JVLA and
\chandra\ positions we find $P_{cc}\approx 0.006$ and $\approx 0.001$,
respectively.  Thus, the sub-arcsecond radio and X-ray positions
enable a robust association with the galaxy, while the XRT position
alone would have led to marginal confidence.  This is essential since
the detection of the host galaxy in the F606W filter limits the
redshift to $z\lesssim 4$, indicating that GRB\,110709B is dark due to
extinction and not a $z\gtrsim 18$ origin.

At $z\sim 2$, the observed F606W band samples the rest-frame UV
emission from the host galaxy.  Based on the observed flux density we
infer a star formation rate of ${\rm SFR}\approx 1$ M$_\odot$
yr$^{-1}$ \citep{ken98}, not corrected for possible galactic-scale
extinction.

\subsubsection{GRB\,111215A}

\begin{figure}
\centering
\includegraphics[totalheight=0.8\textwidth]{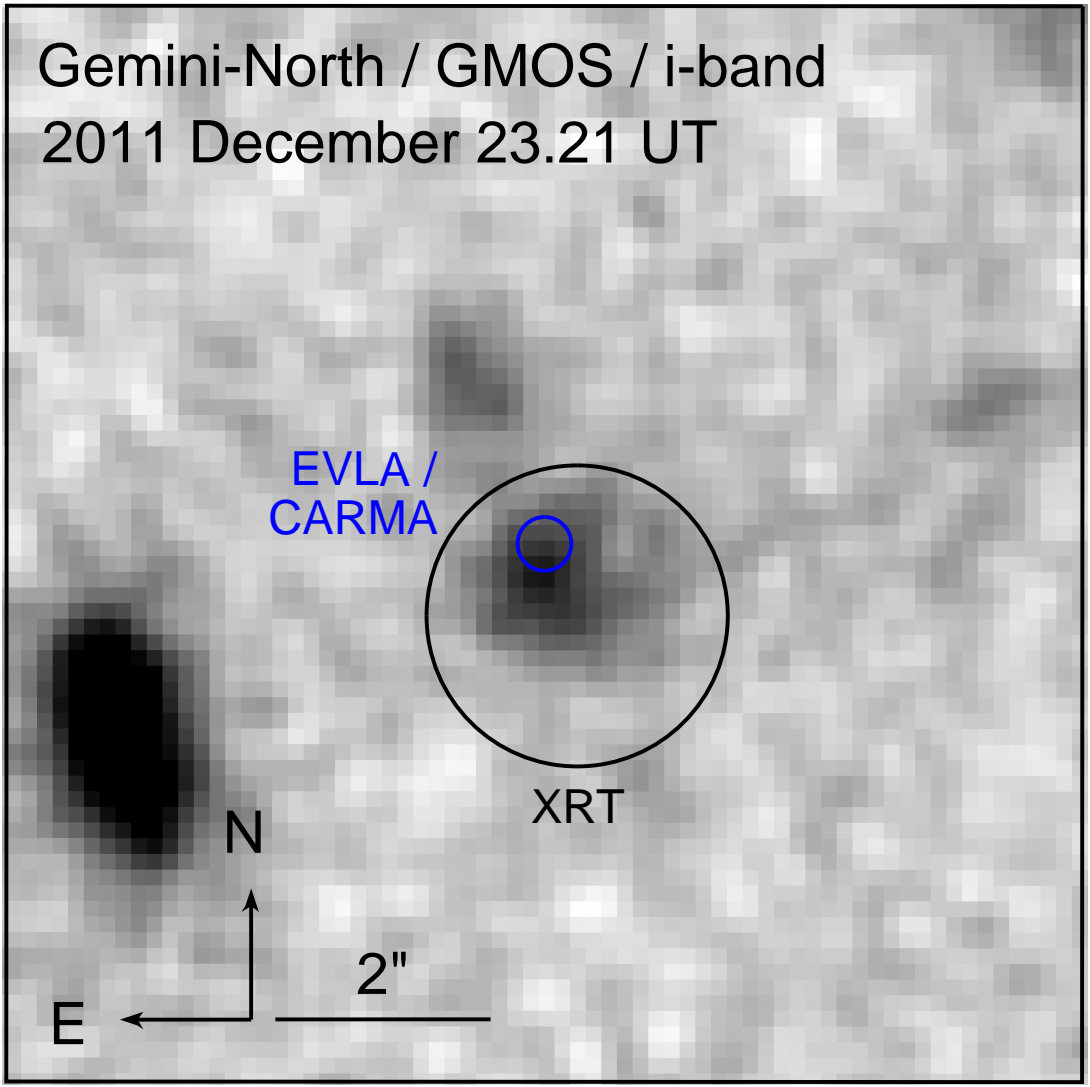}
\caption{Gemini/GMOS $i$-band image of the host galaxy of
GRB\,111215A.  The XRT error circle ($1.4''$ radius; $90\%$
containment) is indicated with the black circle.  The JVLA and CARMA
afterglow positions are shown with a blue circle ($0.13''$; $1\sigma$ systematic)
coinciding with a galaxy that we consider to be the host of
GRB\,111215A.
\label{fig:111215host}}
\end{figure}

We obtained $i$-band observations of GRB\,111215A with GMOS on the
Gemini-North 8-m telescope on 2011 December 23.21 UT ($\delta t\approx
7.6$ d), with a total exposure time of 2160 s in $0.9''$ seeing.
Astrometric matching to the 2MASS catalog using 30 common sources
reveals a galaxy within the enhanced XRT error circle at
RA=\ra{23}{18}{13.317}, Dec=\dec{+32}{29}{38.77}, with an uncertainty
of about $0.16''$ in each coordinate (absolute) and an uncertainty of
about $0.06''$ in the galaxy centroid; see
Figure~\ref{fig:111215host}.  We also obtained imaging observations
with the Low-Resolution Imaging Spectrometer (LRIS; \citealt{Oke+95})
on the Keck I 10-m telescope on 2012 July 15.20 UT ($\delta t = 212.6$
d) in the $g$- and $I$-band filters.

Taking into account the uncertainty in the JVLA radio afterglow
position ($0.06''$ in each coordinate), the resulting overall
uncertainty in the location of the afterglow on the Gemini image is
$0.26''$ radius ($1\sigma$).  Similarly, the uncertainty in the CARMA
mm afterglow position ($0.08''$ in each coordinate) leads to an
overall uncertainty of $0.27''$ radius ($1\sigma$).  The resulting
offset between the host centroid and radio positions is $0.30\pm
0.26''$.

Photometry of the galaxy relative to the SDSS catalog results in a
brightness of $m_g=24.50\pm 0.15$ mag and $m_i=23.64\pm 0.05$ mag
corrected for Galactic extinction.  The probability of chance
coincidence for this galaxy within the XRT error circle is $P_{\rm
cc}\approx 0.01$, with a slightly lower probability of $\approx
7\times 10^{-3}$ using the radio position, dominated by the physical
extent of the galaxy of about $1''$.  Given the low chance coincidence
probability we consider this galaxy to be the host of GRB\,111215A.

To determine the redshift of the host we obtained an 1800~s spectrum
with LRIS, using a $0.7''$ slit with the 400/3400 grism in the blue
arm and the 400/8500 grating in the red arm, providing effectively
continuous wavelength coverage from the atmospheric cutoff to $10280$
\AA\ (with a total throughput within 50\% of peak to about 10030 \AA).
The slit was oriented to cover the host galaxy, as well as the fainter
extended source to the northeast.  The spectra were reduced using
standard techniques implemented in a custom pipeline.

We detect continuum emission from both galaxies on the blue side, and
from the host galaxy on the red side.  We identify only one emission
feature from the nearby galaxy, a marginally resolved line centered at
8019 \AA.  Interpreted as the [\ion{O}{2}]$\lambda 3727$ doublet, this
indicates a redshift\footnotemark\footnotetext{We also examined the
possibility of alternative line identifications and redshifts, but
these would imply the detection of other spectral lines elsewhere that
are not observed.} of $z=1.152$.  We do not identify any emission
lines from the host galaxy.  Given that the galaxy is reasonably
bright in $g$- and $i$-band and is expected to be actively
star-forming, the lack of emission lines indicates that $z\gtrsim 1.8$
(from the non-detection of [\ion{O}{2}]$\lambda 3727$ to 10030 \AA)
and $z\lesssim 2.7$ (from the absence of a break due to the Lyman
limit to 3400 \AA).

At $z\sim 1.8-2.7$, the observed $i$- and $g$-band fluxes trace the
host galaxy rest-frame UV emission.  The observed color of $g-i\approx
0.85\pm 0.20$ mag is indicative of extinction; for example, it is
well-matched to the observed color of Arp\,220, a local ULIRG.  Using
the observed $i$-band flux density, which is less susceptible to
extinction corrections, we infer ${\rm SFR}\approx 15-30$ M$_\odot$
yr$^{-1}$ ($z=1.8-27$), which should be considered as a minimum value
due to the uncertain extinction.

\section{Afterglow Modeling and Rest-Frame Extinction}
\label{sec:model}

GRBs 110709B and 111215A are dark in both the optical and near-IR
bands due to rest-frame dust extinction; a high-redshift origin is
ruled out by the association with optically-detected host galaxies.
It is also of note that for extreme redshifts (z $>$ 18), the $N_{\rm H,int}$ 
implied by the observed excess in the X-ray spectra would be very
high (e.g. $>$ 10$^{24}$ cm$^{-2}$), further supporting the rejection
of an extreme redshift for either event.
In addition to providing accurate localizations and hence secure host
associations, the radio data also allow us to determine the properties
of the bursts and their local environments.  This not only provides a more 
robust measure of the required rest-frame extinction than using 
$\beta_{\rm OX}$ alone (which requires an assumption about the 
location of the synchrotron cooling frequency), but it also allows us 
to compare the explosion properties of dark and optically-bright bursts.

We model the radio and X-ray data for GRBs 110709B and 111215A using
the standard afterglow synchrotron model \citep{gs02,sph99} to: (i)
determine the expected optical and near-IR brightness, and hence the
required level of rest-frame extinction given the observed limits; and
(ii) determine the properties of the bursts and their circumburst
environments.  We follow the standard assumptions of synchrotron
emission from a power-law distribution of electrons
($N(\gamma)\propto\gamma^{-p}$ for $\gamma\ge\gamma_m$) with constant
fractions of the post-shock energy density imparted to the electrons
($\epsilon_e$) and magnetic fields ($\epsilon_B$).  The additional
free parameters of the model are the isotropic-equivalent blast-wave
kinetic energy ($E_{\rm K,iso}$), the circumburst density,
(parametrized as $n$ for a constant density medium: ISM; or as $A$ for
a wind medium with $\rho(r)=Ar^{-2}$), and a jet break time ($t_j$).
To determine the opening angle ($\theta_j$), we use the conversions
from $t_j$ given by \citet{sph99} and \citet{cl00}, with the
appropriate dependence on $E_{\rm K,iso}$ and the circumburst density
($n$ or $A$, respectively).

Using the time evolution of the synchrotron spectrum for both the ISM
and wind density profiles in the pre- and post-jet break phases
\citep{cl00,gs02} we simultaneously fit all X-ray and radio
observations for each burst.  The resulting best-fit parameters are
summarized in Tables~\ref{tab:110709model} and \ref{tab:111215model}
using redshifts of $z=1,2,3,4$ for GRB\,110709B and $z=2$ for
GRB\,111215A.  In both cases we find that the most stringent
constraint on the rest-frame extinction are provided by the observed
$K$-band limits.  We use the SMC extinction curve to determine the
rest-frame extinction, although the Milky Way extinction curve gives
similar results.

For GRB\,110709B we find that the data favor a wind environment, with
a jet break at $t_j\approx 3.1-3.6$ d, corresponding to an opening
angle, $\theta_j$, $\approx 9-23^\circ$; the jet break occurs earlier
and the resulting opening angle is narrower as the redshift increases
from $z=1$ to $z=4$.  The resulting beaming-corrected energies are
$E_K\approx (0.4-1.7)\times 10^{51}$ erg and $E_\gamma\approx
(5.4-10.6)\times 10^{51}$ erg (in the observed $20-5000$ keV fluence
of the initial event).  The circumburst density is characterized by a
mass loss rate of $\dot{M}\approx (4-8)\times 10^{-5}$ M$_\odot$
yr$^{-1}$ for a wind velocity of $v_w=10^3$ km s$^{-1}$ (i.e.,
$A^*\approx 4-8$).  Finally, using the near-IR $K_{\rm s}$-band limit
we find that the required extinction ranges from $A_V^{\rm
host}\gtrsim 10.5$ mag at $z=1$ to $\gtrsim 3.5$ mag at $z=4$.
We note that the best-fit ISM model also requires a high 
density of $\sim 10-100$ cm$^{-3}$ at $z\sim 1-4$.

For GRB\,111215A we again find that a wind environment provides a
better fit, with a jet break at $t_j\approx 12$ d, corresponding to an
opening angle of $\theta_j\approx 24^\circ$.  The resulting
beaming-corrected energies are $E_K\approx 3.7\times 10^{51}$ erg and
$E_\gamma\approx 3.9\times 10^{51}$ erg (in the observed $15-150$ keV
band).  The circumburst density is characterized by a substantial mass
loss rate of $\dot{M}\approx 2\times 10^{-4}$ M$_\odot$ yr$^{-1}$ (for
a wind velocity of $v_w=10^3$ km s$^{-1}$).  Finally, using the
near-IR $K_{\rm s}$-band limit we find a required extinction of
$A_V^{\rm host}\gtrsim 8.5$ mag. We note that the best-fit ISM model 
also requires a density of $\sim 100-300$ cm$^{-3}$.

\section{Discussion}
\label{sec:disc}

\subsection{Extinction and Neutral Hydrogen Column Density}

\begin{figure}
\centering
\includegraphics[totalheight=0.8\textwidth]{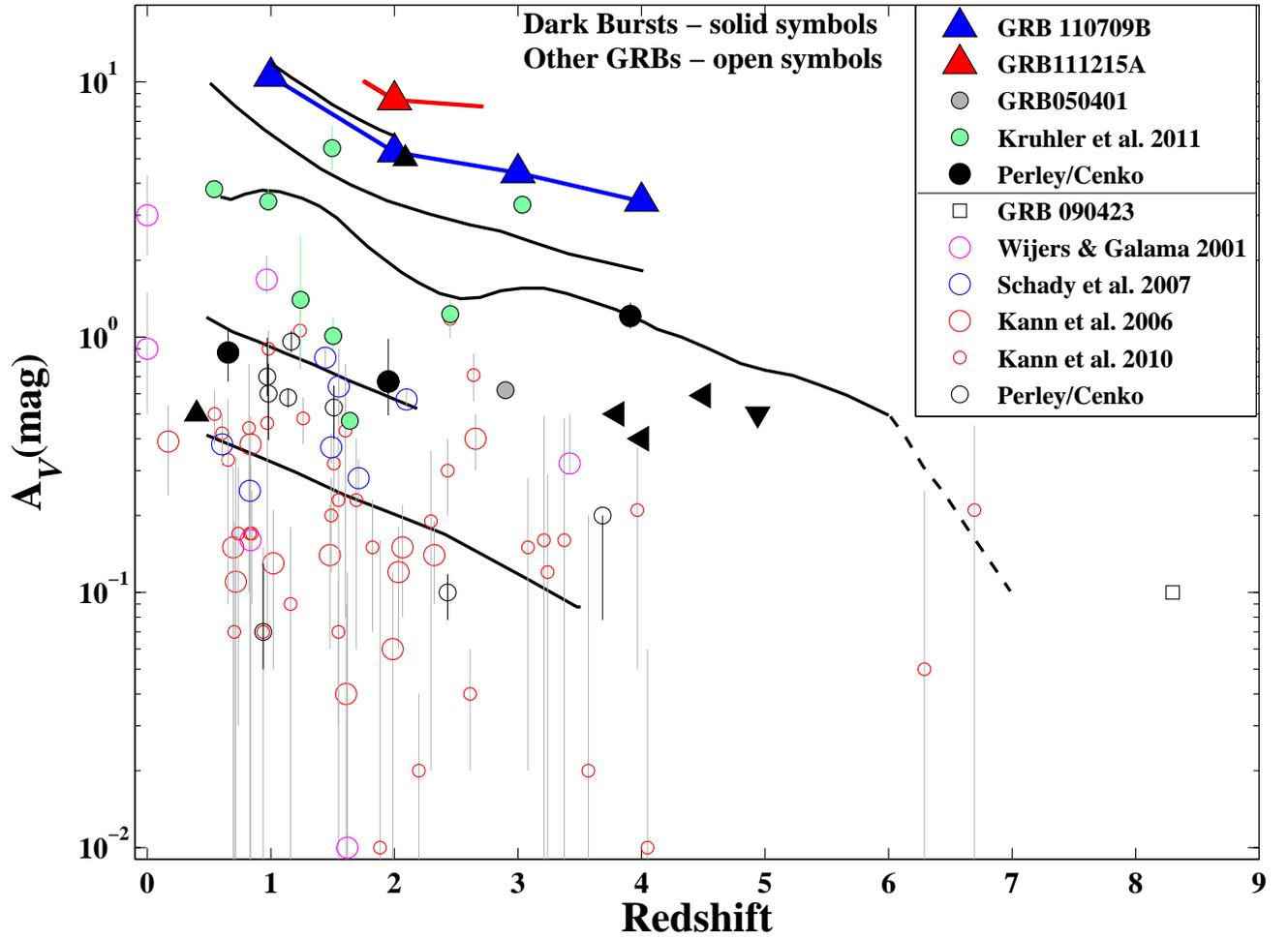}
\caption{Rest-frame extinction ($A_V^{\rm host}$) versus redshift for
GRBs 110709B (blue triangles) and 111215A (red triangle); for both
bursts these are lower limits.  Previous dark bursts are marked with
solid symbols.  Triangles indicate limits on $A_V^{\rm host}$ and/or
redshift.  Allowed values for GRBs without firm redshifts are
indicated by black lines \citep{perley+09}.  Optically-bright GRBs are
marked with open symbols; these bursts generally have $A_V^{\rm
host}\lesssim 1$ mag.  GRBs 110709B and 111215A exhibit some of the
highest extinction values to date.
\label{fig:AVz}}
\end{figure}

The large rest-frame extinction that is required to explain the lack
of optical and near-IR emission from GRBs 110709B and 111215A is
uncommon amongst known GRBs.  In Figure~\ref{fig:AVz} we plot the
minimum values of $A_V^{\rm host}$ for the two bursts as a function of
redshift, along with several comparison samples from the literature 
of optically-bright bursts and previous dark bursts.  In general, 
optically-bright GRBs have inferred extinction values of $\lesssim 1$ mag, 
with a mean of $\sim 0.2-0.3$ mag \citep{kkz06,Schady+07,pcb+09,kgs+11}.  
Dark GRBs span $A_V^{\rm host}\approx 1-6$ mag \citep{pcb+09,kgs+11}, 
with the largest extinction values of $\sim 5-6$ mag inferred for GRBs 061222,
070306, and 070521 \citep{pcb+09,kgs+11}.  In this context, the
extinction values measured here are at the top of the distribution,
$A_V^{\rm host}\gtrsim 5.3$ mag (110709B) and $\gtrsim 8.5$ mag
(111215A) at $z=2$.

\begin{figure}
\centering
\includegraphics[totalheight=0.65\textwidth]{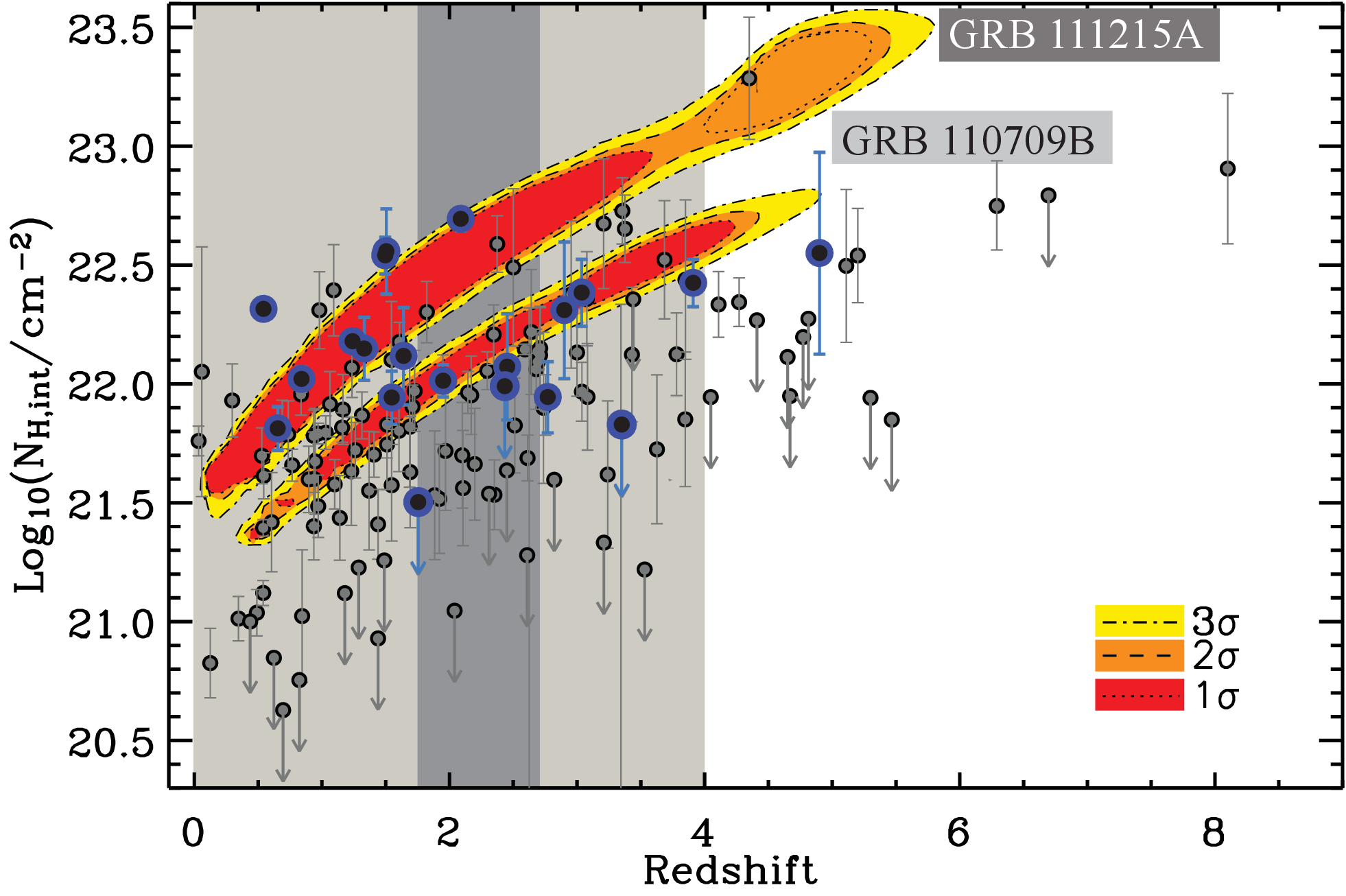}
\caption{Intrinsic neutral hydrogen column density ($N_{\rm H,int}$)
inferred from X-ray observations versus redshift for GRBs 110709B and
111215A (contours).  The allowed redshift ranges for each burst are
marked by the gray shaded regions.  For comparison, we overlay data
for long GRBs in the \swift\ sample up to December 2010
\citep{Margutti+12}; dark bursts are marked by blue/black circles,
while optically-bright events are marked by gray circles.  GRBs
110709B and 111215A clearly lie in the upper portion of the $N_{\rm
H,int}$ distribution of long GRBs, as expected for dark bursts.
\label{fig:NHz}} 
\end{figure}

In Figure~\ref{fig:NHz} we plot the intrinsic neutral hydrogen column
densities, $N_{\rm H,int}$, from our X-ray analysis as a function of redshift, in
comparison with the population of all \swift\ long GRBs with known
redshifts (up to December 2010; \citealt{Margutti+12}).  We find that
GRBs 110709B and 111215A lie at the upper end of the distribution for
long GRBs.  In particular, at a fiducial redshift of $z=2$ they have
${\rm log}(N_{\rm H,int})\approx 22.1$ and $\approx 22.5$,
respectively, in comparison to the median value of all detections and
upper limits of $\langle{\rm log}(N_{\rm H,int})\rangle\approx 21.7$.
This result confirms recent suggestions that dark GRBs have
systematically larger neutral hydrogen columns, generally with ${\rm
log}(N_{\rm H,int})\gtrsim 22$ \citep{perley+09,csm+12}; see
Figure~\ref{fig:NHz}.  Indeed, only a handful of events in the \swift\
sample have values of $N_{\rm H,int}$ that are comparable to that of
GRB\,111215A.  The same conclusion is true for GRB\,110709B if it
resides at the upper end of the allowed redshift distribution ($z\sim
3-4$), although at $z\sim 1$ its neutral hydrogen column is typical of
the overall long GRB population.

\begin{figure}
\centering
\includegraphics[totalheight=0.8\textwidth]{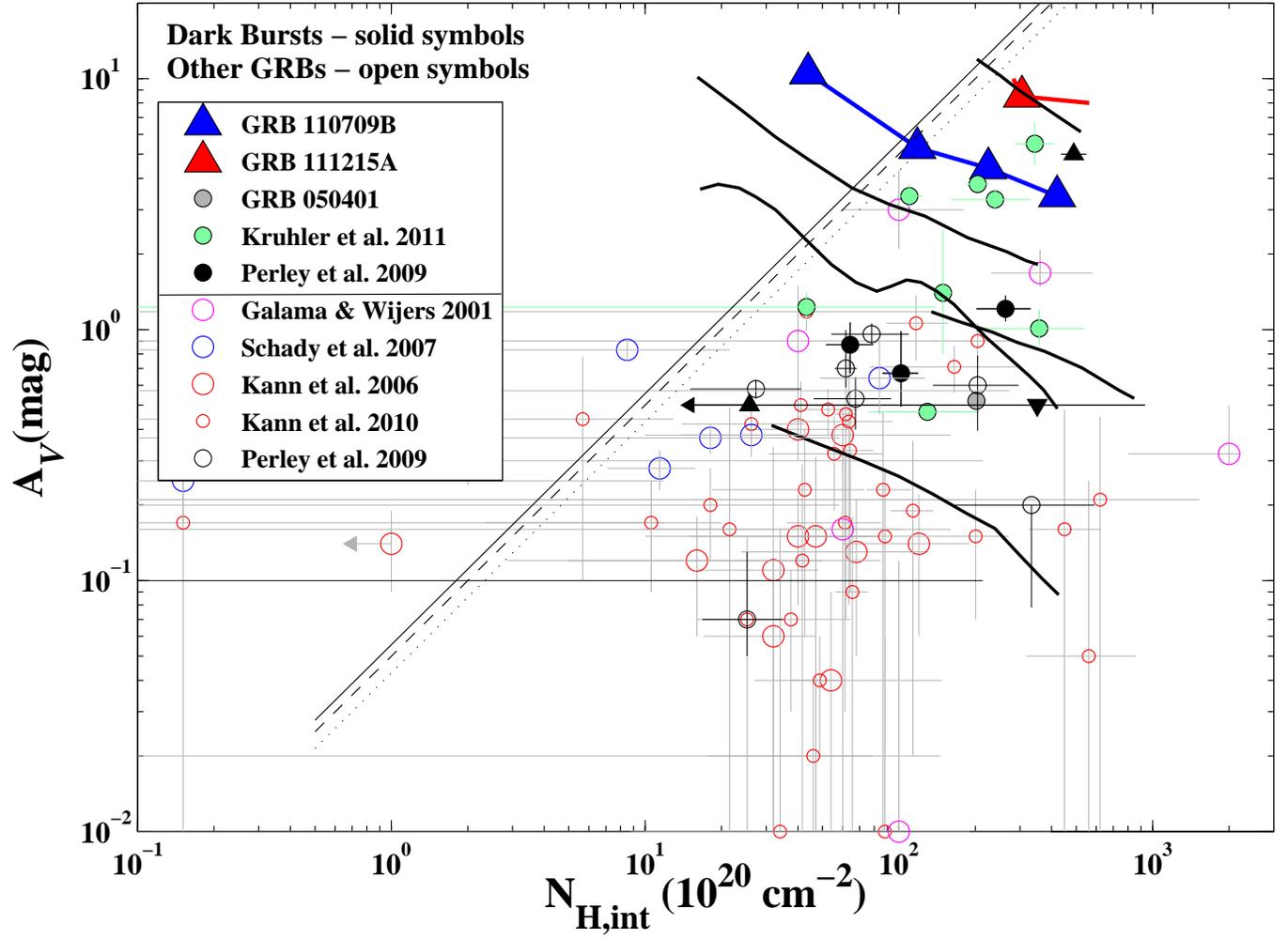}
\caption{Rest-frame extinction ($A_V^{\rm host}$) versus intrinsic
neutral hydrogen column density ($N_{\rm H,int}$) inferred from X-ray
observations for GRBs 110709B (blue triangles) and 111215A (red
triangle).  For comparison, we plot the same sample of bursts shown in
Figure~\ref{fig:AVz}.  The diagonal lines indicate the empirical
$A_V-N_H$ relations for the Milky Way and Magellanic Clouds.  Most
GRBs exhibit lower values of $A_V^{\rm host}$ than expected from the
local relations, but GRBs 110709B and 111215A may be consistent with
these relations.
\label{fig:AVNH}} 
\end{figure}

A correlation between extinction and $N_H$ is known to exist in the
Milky Way and the Magellanic Clouds, with $N_H\approx 2\times
10^{21}\,A_V$ cm$^{-2}$ \citep{ps95,go09}.  As shown in
Figure~\ref{fig:AVNH}, optically-bright GRBs generally have lower
values of $A_V^{\rm host}$ than would be inferred from this relation
(e.g., \citealt{gw01}), with mean values of $N_{\rm H,int}\approx
5\times 10^{21}$ cm$^{-2}$ and $A_V^{\rm host}\approx 0.2-0.3$ mag;
the expected extinction based on the Galactic relation is $A_V^{\rm
host}\approx 2.5$ mag.  Dark GRBs also generally have lower extinction
than expected, with $N_{\rm H,int}\approx (5-50)\times 10^{21}$
cm$^{-2}$ and $A_V^{\rm host} \approx 1-5$ mag, whereas extinctions of
$A_V^{\rm host}\approx 5-25$ mag would be expected.

For the bursts presented here we find potentially different results.
GRB\,111215A agrees with the Galactic relation if it resides at the
lower redshift bound ($z\approx 1.8$), in which case $N_{\rm H,int}
\approx 2.5\times 10^{22}$ cm$^{-2}$ and $A_V^{\rm host}\gtrsim 11$
mag.  At the higher redshift bound ($z\approx 2.7$) we find $N_{\rm
H,int}\approx 5\times 10^{22}$ cm$^{-2}$ and $A_V^{\rm host}\gtrsim 6$
mag (with an expected $A_V^{\rm host}\approx 23$ mag).  This is still
consistent with the Galactic relation since we only place a lower
bound on $A_V^{\rm host}$.  Similarly, GRB\,110709B crosses the
Galactic relation at $z\approx 2$, with $N_{\rm H,int}\approx
1.2\times 10^{22}$ cm$^{-2}$ and $A_V^{\rm host}\gtrsim 5.3$ mag.
However, at $z\lesssim 2$ it has larger extinction than expected; for
example, at $z=1$ we measure $A_V^{\rm host}\gtrsim 10.5$ mag compared
to the expected value of $\approx 2.5$ mag.  At $z\gtrsim 2$ the
situation is similar to GRB\,111215A, namely the minimum value of
$A_V^{\rm host}$ is below the expected Galactic relation, but this is
only a lower limit.  We therefore conclude that the dark bursts with
the largest known extinction values are potentially in line with the
Galactic $N_H-A_V$ relation, although they may also reside below the
expected correlation in line with the bulk of the GRB population.

\subsection{Explosion and Circumburst Properties of Dark Bursts}

A few previous dark GRBs have been localized to sub-arcsecond
precision in the radio.  GRB\,970828 was detected in a single radio
observation with $F_\nu(8.46\,{\rm GHz})=147\pm 33$ $\mu$Jy
($4.5\sigma$), but was not detected in observations only 1 d earlier
and 2 d later \citep{dfk+01}.  This single marginal detection was used
to refine the position from the X-ray region of $10''$ radius, and to
claim an association with a galaxy at $z=0.958$.  The inferred
extinction at this redshift is $A_V^{\rm host}\gtrsim 3.8$ mag.
GRB\,000210 was localized to sub-arcsecond precision with \chandra,
and was also marginally detected in the single epoch in the radio with
$F_\nu(8.46\,{\rm GHz})=93\pm 21$ $\mu$Jy ($4.4\sigma$;
\citealt{pfg+02}).  The associated host galaxy has $z=0.846$, and the
inferred extinction is $A_V^{\rm host}\approx 0.9-3.2$ mag.
GRB\,020819 was detected at high significance in several epochs of
radio observations, leading to an association with a face-on spiral
galaxy at $z=0.410$ \citep{jff+05}.  The required extinction was a
mild $A_V^{\rm host}\approx 0.6-1.5$ mag.  Finally, GRB\,051022 was
localized using radio and millimeter observations to a host galaxy at
$z=0.809$, requiring an extinction of $A_V^{\rm host}\gtrsim 3.5$ mag
\citep{cbm+07,rhw+07}.  Broad-band modeling indicates a
beaming-corrected energy of $\approx 10^{51}$ erg, and a circumburst
density characterized by $\dot{M}\approx 3\times 10^{-7}$ M$_\odot$
yr$^{-1}$ \citep{rhw+07}.

The beaming-corrected energy inferred for GRBs 110709B and 111215A,
$E_\gamma+E_K\approx (7-9)\times 10^{51}$ erg, is not unusual for long
GRBs, although it is somewhat higher than the median value of $\approx
3\times 10^{51}$ erg (e.g., \citealt{pk02,bkp+03,cfh+11}).  On the
other hand, the inferred densities are substantial, with inferred mass
loss rates of about $6\times 10^{-5}$ M$_\odot$ yr$^{-1}$ (110709B)
and $2\times 10^{-4}$ M$_\odot$ yr$^{-1}$ (111215A).  At the radii
appropriate for the jet break times of the two bursts, $r\approx
(1.2-1.4)\times 10^{17}$ cm \citep{cl00}, these mass loss rates
correspond to particle densities of about 100 and 350 cm$^{-3}$,
respectively, higher than typical values for long GRBs, which are $\sim
0.1-30$ cm$^{-3}$ (e.g., \citealt{cl00,pk02}).  If confirmed with
future detailed observations of dark bursts, the high densities may be
indicative of larger mass loss rates for long GRBs in dusty, and hence
metal rich environments.  Such a trend may be expected if the mass
loss is driven by radiation pressure mediated by metal lines
\citep{vk05}.

\subsection{Burst Durations}

The durations of GRB\,110709B ($T_{90}\approx 850$ s) and GRB\,111215A
($T_{90}\approx 800$ s) are extremely long.  To assess whether this type
of unusually long duration correlates with optical/near-IR darkness, we collect all
\swift\ bursts with durations of $\gtrsim 500$~s, excluding nearby
sub-energetic GRBs (e.g., XRF\,060218, GRB\,100316D).  
The sample
includes 9 events\footnotemark\footnotetext{These are GRBs 041219A,
050820A, 060123, 060124, 060929, 091024, 110709B, 111016A, and
111215A.} of which three bursts are dark (060929, 110709B, 111215A),
five have optical and/or near-IR detections with no clear evidence for
extinction, and one event lacks rapid follow-up observations.  Thus,
at least $\sim 1/3$ of these events are dark, similar to the fraction of dark
bursts in the overall long GRB population.  We therefore conclude that
there is no obvious correlation between unusually long duration and 
darkness, with the caveat that the current sample size is small.

\section{Conclusions}
\label{sec:conc}

Using X-ray, optical/near-IR, and radio observations we have
demonstrated that: (i) GRBs 110709B and 111215A are dark bursts; (ii)
they are robustly associated with galaxies at $z\lesssim 4$ (110709B)
and $z\approx 1.8-2.7$ (111215A); (iii) they require unusually large
rest-frame extinction of $A_V^{\rm host}\gtrsim 5.3$ mag (110709B) and
$\gtrsim 8.5$ mag (111215A) at $z=2$; (iv) they exhibit commensurately
large neutral hydrogen column densities in their X-ray spectra,
$N_{\rm H,int}\approx (1-3)\times 10^{22}$ cm$^{-2}$, which at $z\sim
2$ are consistent with the Galactic $N_H-A_V$ relation, unlike the
overall long GRB population; and (v) their circumburst environments on
a sub-parsec scale are shaped by large progenitor mass loss rates of
$\approx (6-20)\times 10^{-5}$ M$_\odot$ yr$^{-1}$.

Radio observations played a critical role in this study for three
reasons.  First, they provided sub-arcsecond positions, which led to a
secure identification of the host galaxies.  This allowed us to
distinguish the extinction scenario from a high-redshift origin.
Second, the combination of radio and X-ray data allowed us to robustly
determine the required extinction, instead of simply assuming an
optical to X-ray spectral index.  Indeed, in cases with only X-ray
data the unknown location of the synchrotron cooling frequency
prevents a unique determination of the extinction.  Finally, the radio
and X-ray data allowed us to determine the burst parameters, including
the geometry, beaming-corrected energy, and the circumburst density.
We find that the energy scale for GRBs 110709B and 111215A is similar
to the overall population of optically-bright GRBs.  However, the
inferred mass loss rates are larger by about an order magnitude
compared to optically-bright bursts, potentially indicating that GRB
progenitors in dusty environments have stronger metal line driven
winds.

The increased sensitivity of the JVLA played an important role in the
study and utilization of the radio afterglow.  ALMA will provide
similar capabilities, with an expected spatial resolution of about
$0.04-5''$ at 100 GHz (depending on configuration), and hence an
assured sub-arcsecond centroiding accuracy for signal-to-noise ratios
of $\gtrsim 10$ even in the most compact configuration.  As
demonstrated here, \chandra\ observations can also provide
sub-arcsecond positions, but such observations may not be available
for all dark GRB candidates (e.g., GRB\,111215A).  In addition,
\chandra\ data will not enhance the ability to determine the required
extinction through broad-band modeling.

The JVLA and ALMA will also allow us to study the host galaxies of GRBs
110709B and 111215A (as well as those of past and future dark bursts)
in greater detail than optical/near-IR studies alone.  In particular,
they will address the question of highly-obscured star formation in
these galaxies, and may even lead to redshift determinations through
the detection of molecular lines.  Existing observations with the VLA and
the MAMBO and SCUBA bolometers (in the small samples published to-date)
have led to only a few detections of 
GRB hosts \citep{bck+03}, with no obvious preference for dark burst hosts
\citep{bbt+03}.  However, the enhanced sensitivity of ALMA will allow
for detections even with star formation rates of tens of M$_\odot$
yr$^{-1}$ at $z\sim 2$.  The combination of detailed host galaxy
properties from rest-frame radio to UV, coupled with measurements of
the local environments of dark bursts through broad-band afterglow
modeling will shed light on the location of the obscuring dust
(interstellar vs.~circumstellar) and the impact of metallicity on GRB
progenitor formation.

\acknowledgments We thank R.~Chary for helpful discussions regarding
obscured star formation in distant galaxies and Y.~Cao for his assistance 
in attaining the optical spectrum of GRB\,111215A while observing at Keck.
The Berger GRB group at Harvard is supported by the National Science 
Foundation under grant AST-1107973, and by NASA/Swift AO7 grant 
NNX12AD69G.  B.A.Z., E.B.,
R.M., W.F., and A.S.~acknowledge partial support of this research
while in residence at the Kavli Institute for Theoretical Physics
under National Science Foundation Grant PHY11-25915.
E.N. acknowledges partial support by an ERC starting grant.
F.O.~acknowledges funding of his Ph.D. through the \emph{Deutscher
Akademischer Austausch-Dienst} (DAAD).  Funding for GROND was
generously granted from the Leibniz Prize to G.~Hasinger (DFG grant HA
1850/28-1).  D.~P.~is supported by grant HST-HF-51296.01-A, provided
by NASA through a Hubble Fellowship grant from the Space Telescope
Science Institute, which is operated by the Association of
Universities for Research in Astronomy, Incorporated, under NASA
contract NAS5-26555.  Support for this work was provided by the
National Aeronautics and Space Administration (NASA) through Chandra
Award Number 09900712 issued by the Chandra X-ray Observatory Center,
which is operated by the SAO for and on behalf of NASA under contract
NAS8-03060.  The JVLA is operated by the
National Radio Astronomy Observatory, a facility of the NSF operated
under cooperative agreement by Associated Universities, Inc.  JVLA
observations were undertaken as part of project numbers 10C-145 and
11B-242.  Support for CARMA construction was derived from the Gordon
and Betty Moore Foundation, the Kenneth T. and Eileen L. Norris
Foundation, the James S. McDonnell Foundation, the Associates of the
California Institute of Technology, the University of Chicago, the
states of California, Illinois, and Maryland, and the NSF. Ongoing
CARMA development and operations are supported by the NSF under a
cooperative agreement, and by the CARMA partner universities.  CARMA
observations were undertaken as part of projects c0773 and cx334.  The
Submillimeter Array is a joint project between the Smithsonian
Astrophysical Observatory (SAO) and the Academia Sinica Institute of
Astronomy and Astrophysics and is funded by the Smithsonian
Institution and the Academia Sinica.  SMA observations were undertaken
as part of project 2011B-S003.  Some observations were obtained at the
Gemini Observatory, which is operated by the Association of
Universities for Research in Astronomy, Inc., under a cooperative
agreement with the NSF on behalf of the Gemini partnership: the NSF
(United States), the Science and Technology Facilities Council (United
Kingdom), the National Research Council (Canada), CONICYT (Chile), the
Australian Research Council (Australia), Minist\'{e}rio da
Ci\^{e}ncia, Tecnologia e Inova\c{c}\~{a}o (Brazil) and Ministerio de
Ciencia, Tecnolog\'{i}a e Innovaci\'{o}n Productiva (Argentina).
Gemini observations were undertaken as part of programs GS-2011A-Q-25
and GN-2011B-Q-10.  Some observations were made with the NASA/ESA
Hubble Space Telescope, obtained at the Space Telescope Science
Institute, which is operated by the Association of Universities for
Research in Astronomy, Inc., under NASA contract NAS 5-26555.  {\it
HST} observations were undertaken as part of program 12378.  This
research has made use of \swift\ data obtained from the High Energy
Astrophysics Science Archive Research Center (HEASARC), provided by
NASA's Goddard Space Flight Center.  This research has made use of the
XRT Data Analysis Software (XRTDAS) developed under the responsibility
of the ASI Science Data Center (ASDC), Italy.

{\it Facilities:} \facility{Swift (XRT)}, \facility{CXO (ACIS-S)},
\facility{Gemini:South (GMOS)}, \facility{Keck:I (LRIS)},
\facility{HST (ACS, WFC3)}, \facility{CARMA}, \facility{SMA},
\facility{JVLA}


\begin{thebibliography}{89}
\expandafter\ifx\csname natexlab\endcsname\relax\def\natexlab#1{#1}\fi

\bibitem[{{Aceituno} {et~al.}(2011){Aceituno}, {Castro-Tirado}, {de Ugarte
  Postigo}, {Tello}, \& {Gorosabel}}]{gcn12688}
{Aceituno}, F., {Castro-Tirado}, A.~J., {de Ugarte Postigo}, A., {Tello},
  J.~C., \& {Gorosabel}, J. 2011, GRB Coordinates Network, 12688, 1

\bibitem[{{Arnaud}(1996){Arnaud}, K.~A.}]{arn96}
{Arnaud}, K.~A. 1996, in Astronomical Society of the Pacific Conference 
 Series, Vol.~101, Astronomical Data Analysis
  Software and Systems V, ed. G.~H. {Jacoby}, \& J. {Barnes}, 17

\bibitem[{{Sault} {et~al.}(1995){Sault}, {Teuben}, \& {Wright}}]{miriad+95}
{Sault}, R.~J., {Teuben}, P.~J., \& {Wright}, M.~C.~H. 1995, in Astronomical
  Society of the Pacific Conference Series, Vol.~77, Astronomical Data Analysis
  Software and Systems IV, ed. R.~A. {Shaw}, H.~E. {Payne}, \& J.~J.~E.
  {Hayes}, 433

\bibitem[{{Barnard} {et~al.}(2003){Barnard}, {Blain}, {Tanvir}, {Natarajan},
  {Smith}, {Wijers}, {Kouveliotou}, {Rol}, {Tilanus}, \& {Vreeswijk}}]{bbt+03}
{Barnard}, V.~E., {et~al.} 2003, \mnras, 338, 1

\bibitem[{{Barthelmy} {et~al.}(2005){Barthelmy}, {Barbier}, {Cummings},
  {Fenimore}, {Gehrels}, {Hullinger}, {Krimm}, {Markwardt}, {Palmer},
  {Parsons}, {Sato}, {Suzuki}, {Takahashi}, {Tashiro}, \& {Tueller}}]{Bart+05}
{Barthelmy}, S.~D., {et~al.} 2005, \ssr, 120, 143

\bibitem[{{Barthelmy} {et~al.}(2011{\natexlab{a}}){Barthelmy}, {Burrows},
  {Cummings}, {Gehrels}, {Gronwall}, {Holland}, {Kennea}, {Markwardt},
  {Palmer}, {Siegel}, {Starling}, \& {Swenson}}]{gcn12124}
---. 2011{\natexlab{a}}, GRB Coordinates Network, 12124, 1

\bibitem[{{Barthelmy} {et~al.}(2011{\natexlab{b}}){Barthelmy}, {Baumgartner},
  {Cummings}, {Fenimore}, {Gehrels}, {Krimm}, {Markwardt}, {Oates}, {Palmer},
  {Sakamoto}, {Sato}, {Stamatikos}, {Tueller}, \& {Ukwatta}}]{gcn12689}
---. 2011{\natexlab{b}}, GRB Coordinates Network, 12689, 1

\bibitem[{{Beardmore} {et~al.}(2011{\natexlab{a}}){Beardmore}, {Evans}, {Goad},
  \& {Osborne}}]{gcn12136}
{Beardmore}, A.~P., {Evans}, P.~A., {Goad}, M.~R., \& {Osborne}, J.~P.
  2011{\natexlab{a}}, GRB Coordinates Network, 12136, 1

\bibitem[{{Beardmore} {et~al.}(2011{\natexlab{b}}){Beardmore}, {Evans}, {Goad},
  \& {Osborne}}]{gcn12690}
---. 2011{\natexlab{b}}, GRB Coordinates Network, 12690, 1

\bibitem[{{Beckwith} {et~al.}(2006){Beckwith}, {Stiavelli}, {Koekemoer},
  {Caldwell}, {Ferguson}, {Hook}, {Lucas}, {Bergeron}, {Corbin}, {Jogee},
  {Panagia}, {Robberto}, {Royle}, {Somerville}, \& {Sosey}}]{bsk+06}
{Beckwith}, S.~V.~W., {et~al.} 2006, \aj, 132, 1729

\bibitem[{{Berger} {et~al.}(2003{\natexlab{a}}){Berger}, {Cowie}, {Kulkarni},
  {Frail}, {Aussel}, \& {Barger}}]{bck+03}
{Berger}, E., {Cowie}, L.~L., {Kulkarni}, S.~R., {Frail}, D.~A., {Aussel}, H.,
  \& {Barger}, A.~J. 2003{\natexlab{a}}, \apj, 588, 99

\bibitem[{{Berger} {et~al.}(2007){Berger}, {Fox}, {Kulkarni}, {Frail}, \&
  {Djorgovski}}]{bfk+07}
{Berger}, E., {Fox}, D.~B., {Kulkarni}, S.~R., {Frail}, D.~A., \& {Djorgovski},
  S.~G. 2007, \apj, 660, 504

\bibitem[{{Berger} {et~al.}(2002){Berger}, {Kulkarni}, {Bloom}, {Price}, {Fox},
  {Frail}, {Axelrod}, {Chevalier}, {Colbert}, {Costa}, {Djorgovski},
  {Frontera}, {Galama}, {Halpern}, {Harrison}, {Holtzman}, {Hurley}, {Kimble},
  {McCarthy}, {Piro}, {Reichart}, {Ricker}, {Sari}, {Schmidt}, {Wheeler},
  {Vanderppek}, \& {Yost}}]{bkb+02}
{Berger}, E., {et~al.} 2002, \apj, 581, 981

\bibitem[{{Berger} {et~al.}(2003{\natexlab{b}}){Berger}, {Kulkarni}, {Pooley},
  {Frail}, {McIntyre}, {Wark}, {Sari}, {Soderberg}, {Fox}, {Yost}, \&
  {Price}}]{bkp+03}
---. 2003{\natexlab{b}}, \nat, 426, 154

\bibitem[{{Bouwens} {et~al.}(2009){Bouwens}, {Illingworth}, {Franx}, {Chary},
  {Meurer}, {Conselice}, {Ford}, {Giavalisco}, \& {van Dokkum}}]{bif+09}
{Bouwens}, R.~J., {et~al.} 2009, \apj, 705, 936

\bibitem[{{Burrows} {et~al.}(2005){Burrows}, {Hill}, {Nousek}, {Kennea},
  {Wells}, {Osborne}, {Abbey}, {Beardmore}, {Mukerjee}, {Short}, {Chincarini},
  {Campana}, {Citterio}, {Moretti}, {Pagani}, {Tagliaferri}, {Giommi},
  {Capalbi}, {Tamburelli}, {Angelini}, {Cusumano}, {Br{\"a}uninger}, {Burkert},
  \& {Hartner}}]{Burrows05}
{Burrows}, D.~N., {et~al.} 2005, \ssr, 120, 165

\bibitem[{{Campana} {et~al.}(2012){Campana}, {Salvaterra}, {Melandri},
  {Vergani}, {Covino}, {D'Avanzo}, {Fugazza}, {Ghisellini}, {Sbarufatti}, \&
  {Tagliaferri}}]{csm+12}
{Campana}, S., {et~al.} 2012, \mnras, 421, 1697

\bibitem[{{Castro-Tirado} {et~al.}(2007){Castro-Tirado}, {Bremer}, {McBreen},
  {Gorosabel}, {Guziy}, {Fakthullin}, {Sokolov}, {Gonz{\'a}lez Delgado},
  {Bihain}, {Pandey}, {Jel{\'{\i}}nek}, {de Ugarte Postigo}, {Misra}, {Sagar},
  {Bama}, {Kamble}, {Anupama}, {Licandro}, {P{\'e}rez-Ram{\'{\i}}rez},
  {Bhattacharya}, {Aceituno}, \& {Neri}}]{cbm+07}
{Castro-Tirado}, A.~J., {et~al.} 2007, \aap, 475, 101

\bibitem[{{Cenko} {et~al.}(2009){Cenko}, {Kelemen}, {Harrison}, {Fox},
  {Kulkarni}, {Kasliwal}, {Ofek}, {Rau}, {Gal-Yam}, {Frail}, \&
  {Moon}}]{Cenko+09}
{Cenko}, S.~B., {et~al.} 2009, \apj, 693, 1484

\bibitem[{{Cenko} {et~al.}(2011){Cenko}, {Frail}, {Harrison}, {Haislip},
  {Reichart}, {Butler}, {Cobb}, {Cucchiara}, {Berger}, {Bloom}, {Chandra},
  {Fox}, {Perley}, {Prochaska}, {Filippenko}, {Glazebrook}, {Ivarsen},
  {Kasliwal}, {Kulkarni}, {LaCluyze}, {Lopez}, {Morgan}, {Pettini}, \&
  {Rana}}]{cfh+11}
---. 2011, \apj, 732, 29

\bibitem[{{Chevalier} \& {Li}(2000)}]{cl00}
{Chevalier}, R.~A., \& {Li}, Z.-Y. 2000, \apj, 536, 195

\bibitem[{{Cucchiara} {et~al.}(2011){Cucchiara}, {Levan}, {Fox}, {Tanvir},
  {Ukwatta}, {Berger}, {Kr{\"u}hler}, {K{\"u}pc{\"u} Yolda{\c s}}, {Wu},
  {Toma}, {Greiner}, {Olivares}, {Rowlinson}, {Amati}, {Sakamoto}, {Roth},
  {Stephens}, {Fritz}, {Fynbo}, {Hjorth}, {Malesani}, {Jakobsson}, {Wiersema},
  {O'Brien}, {Soderberg}, {Foley}, {Fruchter}, {Rhoads}, {Rutledge}, {Schmidt},
  {Dopita}, {Podsiadlowski}, {Willingale}, {Wolf}, {Kulkarni}, \&
  {D'Avanzo}}]{clf+11}
{Cucchiara}, A., {et~al.} 2011, \apj, 736

\bibitem[{{Cummings} {et~al.}(2011a){Cummings}, {Barthelmy}, {Burrows},
  {Gronwall}, {Holland}, {Kennea}, {Markwardt}, {Palmer}, {Siegel}, {Starling},
  \& {Swenson}}]{gcn12122}
{Cummings}, J.~R., {et~al.} 2011, GRB Coordinates Network, 12122, 1

\bibitem[{{Cummings} {et~al.}(2011b){Cummings}, {Barthelmy}, {Baumgartner},
{Fenimore},{Gehrels},{Krimm},{Markwardt},{Palmer},{Parsons},{Sakamoto},
{Stamatikos},{Tueller}, \& {Ukwatta}}]{gcn12144}
---. 2011, GRB Coordinates Network, 12144, 1

\bibitem[{{de Ugarte Postigo} {et~al.}(2011){de Ugarte Postigo}, {Lundgren},
  {De Breuck}, {Siringo}, {Parra}, \& {Agurto}}]{gcn12151}
{de Ugarte Postigo}, A., {Lundgren}, A., {De Breuck}, C., {Siringo}, G.,
  {Parra}, R., \& {Agurto}, C. 2011, GRB Coordinates Network, 12151, 1

\bibitem[{{Djorgovski} {et~al.}(2001){Djorgovski}, {Frail}, {Kulkarni},
  {Bloom}, {Odewahn}, \& {Diercks}}]{dfk+01}
{Djorgovski}, S.~G., {Frail}, D.~A., {Kulkarni}, S.~R., {Bloom}, J.~S.,
  {Odewahn}, S.~C., \& {Diercks}, A. 2001, \apj, 562, 654

\bibitem[{{Djorgovski} {et~al.}(1998){Djorgovski}, {Kulkarni}, {Bloom},
  {Goodrich}, {Frail}, {Piro}, \& {Palazzi}}]{dkb+98}
{Djorgovski}, S.~G., {Kulkarni}, S.~R., {Bloom}, J.~S., {Goodrich}, R.,
  {Frail}, D.~A., {Piro}, L., \& {Palazzi}, E. 1998, \apjl, 508, L17

\bibitem[{{D$^{\prime}$Avanzo}(2011)}]{gcn12695}
{D$^{\prime}$Avanzo}, A.~e.~a. 2011, GRB Coordinates Network, 12695, 1

\bibitem[{{Fruchter} \& {Hook}(2002)}]{fh02}
{Fruchter}, A.~S., \& {Hook}, R.~N. 2002, \pasp, 114, 144

\bibitem[{{Fruchter} {et~al.}(2006){Fruchter}, {Levan}, {Strolger},
  {Vreeswijk}, {Thorsett}, {Bersier}, {Burud}, {Castro Cer{\'o}n},
  {Castro-Tirado}, {Conselice}, {Dahlen}, {Ferguson}, {Fynbo}, {Garnavich},
  {Gibbons}, {Gorosabel}, {Gull}, {Hjorth}, {Holland}, {Kouveliotou}, {Levay},
  {Livio}, {Metzger}, {Nugent}, {Petro}, {Pian}, {Rhoads}, {Riess}, {Sahu},
  {Smette}, {Tanvir}, {Wijers}, \& {Woosley}}]{fls+06}
{Fruchter}, A.~S., {et~al.} 2006, \nat, 441, 463

\bibitem[{{Fynbo} {et~al.}(2001){Fynbo}, {Jensen}, {Gorosabel}, {Hjorth},
  {Pedersen}, {M{\o}ller}, {Abbott}, {Castro-Tirado}, {Delgado}, {Greiner},
  {Henden}, {Magazz{\`u}}, {Masetti}, {Merlino}, {Masegosa}, {{\O}stensen},
  {Palazzi}, {Pian}, {Schwarz}, {Cline}, {Guidorzi}, {Goldsten}, {Hurley},
  {Mazets}, {McClanahan}, {Montanari}, {Starr}, \& {Trombka}}]{fjg+01}
{Fynbo}, J.~U., {et~al.} 2001, \aap, 369, 373

\bibitem[{{Galama} \& {Wijers}(2001)}]{gw01}
{Galama}, T.~J., \& {Wijers}, R.~A.~M.~J. 2001, \apjl, 549, L209

\bibitem[{{Golenetskii} {et~al.}(2011){Golenetskii}, {Aptekar}, {Frederiks},
  {Mazets}, {Pal'Shin}, {Oleynik}, {Ulanov}, {Svinkin}, \&
  {Cline}}]{Golentskii+11}
{Golenetskii}, S., {et~al.} 2011, GRB Coordinates Network, 12135, 1

\bibitem[{{Gorbovskoy} {et~al.}(2011){Gorbovskoy}, {Lipunov}, {Kornilov},
  {Kuvshinov}, {Belinski}, {Tyurina}, {Shatskiy}, {Balanutsa}, {Chazov},
  {Kuznetsov}, {Zimnukhov}, {Kornilov}, {Sankovich}, {Shurpakov}, {Ivanov},
  {Poleshchuk}, {Yazev}, {Budnev}, {Gres}, {Chuvalaev}, {Konstantinov},
  {Sinykov}, {Yurkov}, {Sergienko}, {Varda}, {Tlatov}, {Parhomenko},
  {Dormidontov}, {Sennik}, {Krushinski}, {Zalozhnich}, {Kopytova}, \&
  {Popov}}]{gcn12687}
{Gorbovskoy}, E., {et~al.} 2011, GRB Coordinates Network, 12687, 1

\bibitem[{{Granot} \& {Sari}(2002)}]{gs02}
{Granot}, J., \& {Sari}, R. 2002, \apj, 568, 820

\bibitem[{{Greiner} {et~al.}(2007){Greiner}, {Bornemann}, {Clemens}, {Deuter},
  {Hasinger}, {Honsberg}, {Huber}, {Huber}, {Krauss}, {Kr{\"u}hler},
  {K{\"u}pc{\"u} Yoldas}, {Mayer-Hasselwander}, {Mican}, {Primak}, {Schrey},
  {Steiner}, {Szokoly}, {Th{\"o}ne}, {Yoldas}, {Klose}, {Laux}, \&
  {Winkler}}]{Greiner07}
{Greiner}, J., {et~al.} 2007, The Messenger, 130, 12

\bibitem[{{Greiner} {et~al.}(2008){Greiner}, {Bornemann}, {Clemens}, {Deuter},
  {Hasinger}, {Honsberg}, {Huber}, {Huber}, {Krauss}, {Kr{\"u}hler},
  {K{\"u}pc{\"u} Yolda{\c s}}, {Mayer-Hasselwander}, {Mican}, {Primak},
  {Schrey}, {Steiner}, {Szokoly}, {Th{\"o}ne}, {Yolda{\c s}}, {Klose}, {Laux},
  \& {Winkler}}]{Greiner08}
---. 2008, \pasp, 120, 405

\bibitem[{{Greiner} {et~al.}(2011){Greiner}, {Kr{\"u}hler}, {Klose}, {Afonso},
  {Clemens}, {Filgas}, {Hartmann}, {K{\"u}pc{\"u} Yolda{\c s}}, {Nardini},
  {Olivares E.}, {Rau}, {Rossi}, {Schady}, \& {Updike}}]{gkk+11}
---. 2011, \aap, 526, A30

\bibitem[{{Greisen}(2003)}]{greisen03}
{Greisen}, E.~W. 2003, Information Handling in Astronomy - Historical Vistas,
  285, 109

\bibitem[{{Groot} {et~al.}(1998){Groot},{Galama},{van Paradijs},{Kouveliotou},
{Wijers},{Bloom},{Tanvir},{Vanderspek},{Greiner},{Castro-Tirado},{Gorosabel},
{von Hippel},{Lehnert},{Kuijken},{Hoekstra},{Metcalfe},{Howk},{Conselice},
{Telting},{Rutten},{Rhoads},{Cole},{Pisano},{Naber}, \& {Schwarz}}]{ggp+98}
{Groot}, P.~J. {et~al.} 1998, \apj, 493, L27

\bibitem[{{G{\"u}ver} \& {{\"O}zel}(2009)}]{go09}
{G{\"u}ver}, T., \& {{\"O}zel}, F. 2009, \mnras, 400, 2050

\bibitem[{{Haislip} {et~al.}(2006){Haislip}, {Nysewander}, {Reichart}, {Levan},
  {Tanvir}, {Cenko}, {Fox}, {Price}, {Castro-Tirado}, {Gorosabel}, {Evans},
  {Figueredo}, {MacLeod}, {Kirschbrown}, {Jelinek}, {Guziy}, {Postigo},
  {Cypriano}, {Lacluyze}, {Graham}, {Priddey}, {Chapman}, {Rhoads}, {Fruchter},
  {Lamb}, {Kouveliotou}, {Wijers}, {Bayliss}, {Schmidt}, {Soderberg},
  {Kulkarni}, {Harrison}, {Moon}, {Gal-Yam}, {Kasliwal}, {Hudec}, {Vitek},
  {Kubanek}, {Crain}, {Foster}, {Clemens}, {Bartelme}, {Canterna}, {Hartmann},
  {Henden}, {Klose}, {Park}, {Williams}, {Rol}, {O'Brien}, {Bersier}, {Prada},
  {Pizarro}, {Maturana}, {Ugarte}, {Alvarez}, {Fernandez}, {Jarvis}, {Moles},
  {Alfaro}, {Ivarsen}, {Kumar}, {Mack}, {Zdarowicz}, {Gehrels}, {Barthelmy}, \&
  {Burrows}}]{hnr+06}
{Haislip}, J.~B., {et~al.} 2006, \nat, 440, 181

\bibitem[{{Ho} {et~al.}(2004){Ho}, {Moran}, \& {Lo}}]{Ho+04}
{Ho}, P.~T.~P., {Moran}, J.~M., \& {Lo}, K.~Y. 2004, \apjl, 616, L1

\bibitem[{{Hogg} {et~al.}(1997){Hogg}, {Pahre}, {McCarthy}, {Cohen},
  {Blandford}, {Smail}, \& {Soifer}}]{hpm+97}
{Hogg}, D.~W., {Pahre}, M.~A., {McCarthy}, J.~K., {Cohen}, J.~G., {Blandford},
  R., {Smail}, I., \& {Soifer}, B.~T. 1997, \mnras, 288, 404

\bibitem[{{Hook} {et~al.}(2004){Hook}, {J{\o}rgensen}, {Allington-Smith},
  {Davies}, {Metcalfe}, {Murowinski}, \& {Crampton}}]{Hook+2004}
{Hook}, I.~M., {J{\o}rgensen}, I., {Allington-Smith}, J.~R., {Davies}, R.~L.,
  {Metcalfe}, N., {Murowinski}, R.~G., \& {Crampton}, D. 2004, \pasp, 116, 425

\bibitem[{{Jakobsson} {et~al.}(2004){Jakobsson}, {Hjorth}, {Fynbo}, {Watson},
  {Pedersen}, {Bj{\"o}rnsson}, \& {Gorosabel}}]{jakobsson+04}
{Jakobsson}, P., {Hjorth}, J., {Fynbo}, J.~P.~U., {Watson}, D., {Pedersen}, K.,
  {Bj{\"o}rnsson}, G., \& {Gorosabel}, J. 2004, \apjl, 617, L21

\bibitem[{{Jakobsson} {et~al.}(2005){Jakobsson}, {Frail}, {Fox}, {Moon},
  {Price}, {Kulkarni}, {Fynbo}, {Hjorth}, {Berger}, {McNaught}, \&
  {Dahle}}]{jff+05}
{Jakobsson}, P., {et~al.} 2005, \apj, 629, 45

\bibitem[{{Kalberla} {et~al.}(2005){Kalberla}, {Burton}, {Hartmann}, {Arnal},
  {Bajaja}, {Morras}, \& {P{\"o}ppel}}]{Kalberla05}
{Kalberla}, P.~M.~W., {Burton}, W.~B., {Hartmann}, D., {Arnal}, E.~M.,
  {Bajaja}, E., {Morras}, R., \& {P{\"o}ppel}, W.~G.~L. 2005, \aap, 440, 775

\bibitem[{{Kann} {et~al.}(2006){Kann}, {Klose}, \& {Zeh}}]{kkz06}
{Kann}, D.~A., {Klose}, S., \& {Zeh}, A. 2006, \apj, 641, 993

\bibitem[{{Kennicutt}(1998)}]{ken98}
{Kennicutt}, Jr., R.~C. 1998, \araa, 36, 189

\bibitem[{{Kimble} {et~al.}(2008){Kimble}, {MacKenty}, {O'Connell}, \&
  {Townsend}}]{Kimble+2008}
{Kimble}, R.~A., {MacKenty}, J.~W., {O'Connell}, R.~W., \& {Townsend}, J.~A.
  2008, in Society of Photo-Optical Instrumentation Engineers (SPIE) Conference
  Series, Vol. 7010, Society of Photo-Optical Instrumentation Engineers (SPIE)
  Conference Series

\bibitem[{{Koekemoer} {et~al.}(2006){Koekemoer}, {Fruchter}, {Hook}, {Hack}, \&
  {Hanley}}]{kfh+06}
{Koekemoer}, A.~M., {Fruchter}, A.~S., {Hook}, R.~N., {Hack}, W., \& {Hanley},
  C. 2006, in The 2005 HST Calibration Workshop: Hubble After the Transition to
  Two-Gyro Mode, ed. A.~M. {Koekemoer}, P.~{Goudfrooij}, \& L.~L. {Dressel},
  423

\bibitem[{{Kr{\"u}hler} {et~al.}(2008){Kr{\"u}hler}, {K{\"u}pc{\"u} Yolda{\c
  s}}, {Greiner}, {Clemens}, {McBreen}, {Primak}, {Savaglio}, {Yolda{\c s}},
  {Szokoly}, \& {Klose}}]{Kruehler08}
{Kr{\"u}hler}, T., {et~al.} 2008, \apj, 685, 376

\bibitem[{{Kr{\"u}hler} {et~al.}(2011){Kr{\"u}hler}, {Greiner}, {Schady},
  {Savaglio}, {Afonso}, {Clemens}, {Elliott}, {Filgas}, {Gruber}, {Kann},
  {Klose}, {K{\"u}pc{\"u}-Yolda{\c s}}, {McBreen}, {Olivares}, {Pierini},
  {Rau}, {Rossi}, {Nardini}, {Nicuesa Guelbenzu}, {Sudilovsky}, \&
  {Updike}}]{kgs+11}
---. 2011, \aap, 534, A108

\bibitem[{{Levan} {et~al.}(2006){Levan}, {Fruchter},{Rhoads},{Mobasher},
  {Tanvir},{Gorosabel},{Rol},{Kouveliotou},{Dell Antonio},{Merrill},{Bergeron},
  {Castro Cer{\'o}n},{Masetti},{Vreeswijk},{Antonelli},{Bersier},{Castro-Tirado},
  {Fynbo},{Garnavich},{Holland},{Hjorth},{Nugent},{Pian},{Smette},{Thomsen},
  {Thorsett}, \& {Wijers}}]{lfr+06}
 {Levan}, A., {et~al.} 2006, \apj, 647, 471

\bibitem[{{Levesque} {et~al.}(2010){Levesque}, {Kewley}, {Graham}, \&
  {Fruchter}}]{lkg+10}
{Levesque}, E.~M., {Kewley}, L.~J., {Graham}, J.~F., \& {Fruchter}, A.~S. 2010,
  \apjl, 712, L26

\bibitem[{{Margutti} {et~al.}(2012){Margutti}, {Zaninoni}, {Bernardini},
  {Chincarini}, \& {for the Swift-XRT team}}]{Margutti+12}
{Margutti}, R., {Zaninoni}, E., {Bernardini}, M.~G., {Chincarini}, G., \& {for
  the Swift-XRT team}. 2012, ArXiv e-prints

\bibitem[{{Margutti} {et~al.}(2010){Margutti}, {Genet}, {Granot}, {Barniol
  Duran}, {Guidorzi}, {Chincarini}, {Mao}, {Schady}, {Sakamoto}, {Miller},
  {Olofsson}, {Bloom}, {Evans}, {Fynbo}, {Malesani}, {Moretti}, {Pasotti},
  {Starr}, {Burrows}, {Barthelmy}, {Roming}, \& {Gehrels}}]{Margutti10}
{Margutti}, R., {et~al.} 2010, \mnras, 402, 46

\bibitem[{{Melandri} {et~al.}(2008){Melandri}, {Mundell}, {Kobayashi},
  {Guidorzi}, {Gomboc}, {Steele}, {Smith}, {Bersier}, {Mottram}, {Carter},
  {Bode}, {O'Brien}, {Tanvir}, {Rol}, \& {Chapman}}]{mmk+08}
{Melandri}, A., {et~al.} 2008, \apj, 686, 1209

\bibitem[{{Melandri} {et~al.}(2012){Melandri}, {Sbarufatti}, {D'Avanzo},
  {Salvaterra}, {Campana}, {Covino}, {Vergani}, {Nava}, {Ghisellini},
  {Ghirlanda}, {Fugazza}, {Mangano}, {Capalbi}, \& {Tagliaferri}}]{Melandri+12}
---. 2012, \mnras, 421, 1265

\bibitem[{{Murphy} {et~al.}(2011){Murphy}, {Chary}, {Dickinson}, {Pope},
  {Frayer}, \& {Lin}}]{mcd+11}
{Murphy}, E.~J., {Chary}, R.-R., {Dickinson}, M., {Pope}, A., {Frayer}, D.~T.,
  \& {Lin}, L. 2011, \apj, 732, 126

\bibitem[{{Oates}(2011)}]{gcn12693}
{Oates}, S.~R. 2011, GRB Coordinates Network, 12693, 1

\bibitem[{{Oates} {et~al.}(2011){Oates}, {Barthelmy}, {Baumgartner}, {Burrows},
  {Cummings}, {D'Elia}, {Guidorzi}, T., {Krimm}, {Kuin}, {Mangano}, {Marshall},
  {Mountford}, {O'Brien}, {Osborne}, {Pagani}, {Page}, {Palmer}, {Romano},
  {Siegel}, {Starling}, {Ukwatta}, \& {Zhang}}]{gcn12681}
{Oates}, S.~R., {et~al.} 2011, GRB Coordinates Network, 12681, 1

\bibitem[{{Oke} {et~al.}(1995){Oke}, {Cohen}, {Carr}, {Cromer}, {Dingizian},
  {Harris}, {Labrecque}, {Lucinio}, {Schaal}, {Epps}, \& {Miller}}]{Oke+95}
{Oke}, J.~B., {et~al.} 1995, \pasp, 107, 375

\bibitem[{{Panaitescu} \& {Kumar}(2002)}]{pk02}
{Panaitescu}, A., \& {Kumar}, P. 2002, \apj, 571, 779

\bibitem[{{Pandey}(2011)}]{gcn12686}
{Pandey}, S.~B. e.~a. 2011, GRB Coordinates Network, 12686, 1

\bibitem[{{Perley} {et~al.}(2009{\natexlab{a}}){Perley}, {Cenko}, {Bloom},
  {Chen}, {Butler}, {Kocevski}, {Prochaska}, {Brodwin}, {Glazebrook},
  {Kasliwal}, {Kulkarni}, {Lopez}, {Ofek}, {Pettini}, {Soderberg}, \&
  {Starr}}]{perley+09}
{Perley}, D.~A., {et~al.} 2009{\natexlab{a}}, \aj, 138, 1690

\bibitem[{{Perley} {et~al.}(2009{\natexlab{b}}){Perley}, {Cenko}, {Bloom},
  {Chen}, {Butler}, {Kocevski}, {Prochaska}, {Brodwin}, {Glazebrook},
  {Kasliwal}, {Kulkarni}, {Lopez}, {Ofek}, {Pettini}, {Soderberg}, \&
  {Starr}}]{pcb+09}
---. 2009{\natexlab{b}}, \aj, 138, 1690

\bibitem[{{Perley} {et~al.}(2011{\natexlab{a}}){Perley}, {Morgan}, {Updike},
  {Yuan}, {Akerlof}, {Miller}, {Bloom}, {Cenko}, {Li}, {Filippenko},
  {Prochaska}, {Kann}, {Tanvir}, {Levan}, {Butler}, {Christian}, {Hartmann},
  {Milne}, {Rykoff}, {Rujopakarn}, {Wheeler}, \& {Williams}}]{pmu+11}
---. 2011{\natexlab{a}}, \aj, 141, 36

\bibitem[{{Perley} {et~al.}(2011{\natexlab{b}}){Perley}, {Chandler}, {Butler},
  \& {Wrobel}}]{rperley+11}
{Perley}, R.~A., {Chandler}, C.~J., {Butler}, B.~J., \& {Wrobel}, J.~M.
  2011{\natexlab{b}}, \apjl, 739, L1

\bibitem[{{Piro} {et~al.}(2002){Piro}, {Frail}, {Gorosabel}, {Garmire},
  {Soffitta}, {Amati}, {Andersen}, {Antonelli}, {Berger}, {Frontera}, {Fynbo},
  {Gandolfi}, {Garcia}, {Hjorth}, {in 't Zand}, {Jensen}, {Masetti},
  {M{\o}ller}, {Pedersen}, {Pian}, \& {Wieringa}}]{pfg+02}
{Piro}, L., {et~al.} 2002, \apj, 577, 680

\bibitem[{{Predehl} \& {Schmitt}(1995)}]{ps95}
{Predehl}, P., \& {Schmitt}, J.~H.~M.~M. 1995, \aap, 293, 889

\bibitem[{{Reddy} \& {Steidel}(2009)}]{rs09}
{Reddy}, N.~A., \& {Steidel}, C.~C. 2009, \apj, 692, 778

\bibitem[{{Rol} {et~al.}(2007){Rol}, {van der Horst}, {Wiersema}, {Patel},
  {Levan}, {Nysewander}, {Kouveliotou}, {Wijers}, {Tanvir}, {Reichart},
  {Fruchter}, {Graham}, {Ovaldsen}, {Jaunsen}, {Jonker}, {van Ham}, {Hjorth},
  {Starling}, {O'Brien}, {Fynbo}, {Burrows}, \& {Strom}}]{rhw+07}
{Rol}, E., {et~al.} 2007, \apj, 669, 1098

\bibitem[{{Roming} {et~al.}(2005){Roming}, {Kennedy}, {Mason}, {Nousek}, {Ahr},
  {Bingham}, {Broos}, {Carter}, {Hancock}, {Huckle}, {Hunsberger}, {Kawakami},
  {Killough}, {Koch}, {McLelland}, {Smith}, {Smith}, {Soto}, {Boyd},
  {Breeveld}, {Holland}, {Ivanushkina}, {Pryzby}, {Still}, \&
  {Stock}}]{Roming05}
{Roming}, P.~W.~A., {et~al.} 2005, \ssr, 120, 95

\bibitem[{{Rumyantsev}(2011)}]{gcn12703}
{Rumyantsev}, N.~e.~a. 2011, GRB Coordinates Network, 12703, 1

\bibitem[{{Salvaterra} {et~al.}(2009){Salvaterra}, {Della Valle}, {Campana},
  {Chincarini}, {Covino}, {D'Avanzo}, {Fern{\'a}ndez-Soto}, {Guidorzi},
  {Mannucci}, {Margutti}, {Th{\"o}ne}, {Antonelli}, {Barthelmy}, {de Pasquale},
  {D'Elia}, {Fiore}, {Fugazza}, {Hunt}, {Maiorano}, {Marinoni}, {Marshall},
  {Molinari}, {Nousek}, {Pian}, {Racusin}, {Stella}, {Amati}, {Andreuzzi},
  {Cusumano}, {Fenimore}, {Ferrero}, {Giommi}, {Guetta}, {Holland}, {Hurley},
  {Israel}, {Mao}, {Markwardt}, {Masetti}, {Pagani}, {Palazzi}, {Palmer},
  {Piranomonte}, {Tagliaferri}, \& {Testa}}]{sdc+09}
{Salvaterra}, R., {et~al.} 2009, \nat, 461, 1258

\bibitem[{{Sari} {et~al.}(1999){Sari}, {Piran}, \& {Halpern}}]{sph99}
{Sari}, R., {Piran}, T., \& {Halpern}, J.~P. 1999, \apjl, 519, L17

\bibitem[{{Sari} {et~al.}(1998){Sari}, {Piran}, \& {Narayan}}]{spn98}
{Sari}, R., {Piran}, T., \& {Narayan}, R. 1998, \apjl, 497, L17

\bibitem[{{Sault} {et~al.}(1995){Sault}, {Teuben}, \& {Wright}}]{miriad+95}
{Sault}, R.~J., {Teuben}, P.~J., \& {Wright}, M.~C.~H. 1995, in Astronomical
  Society of the Pacific Conference Series, Vol.~77, Astronomical Data Analysis
  Software and Systems IV, ed. R.~A. {Shaw}, H.~E. {Payne}, \& J.~J.~E.
  {Hayes}, 433

\bibitem[{{Sault} {et~al.}(2011){Sault}, {Teuben}, \& {Wright}}]{miriad+11}
{Sault}, R.~J., {Teuben}, P.~J., \& {Wright}, M.~C.~H. 2011, in Astrophysics
  Source Code Library, record ascl:1106.007, 6007

\bibitem[{{Schady} {et~al.}(2007){Schady}, {Mason}, {Page}, {de Pasquale},
  {Morris}, {Romano}, {Roming}, {Immler}, \& {vanden Berk}}]{Schady+07}
{Schady}, P., {et~al.} 2007, \mnras, 377, 273

\bibitem[{{Schlafly} \& {Finkbeiner}(2011)}]{sf11}
{Schlafly}, E.~F., \& {Finkbeiner}, D.~P. 2011, \apj, 737, 103

\bibitem[{{Siringo} {et~al.}(2009){Siringo}, {Kreysa}, {Kov{\'a}cs},
  {Schuller}, {Wei{\ss}}, {Esch}, {Gem{\"u}nd}, {Jethava}, {Lundershausen},
  {Colin}, {G{\"u}sten}, {Menten}, {Beelen}, {Bertoldi}, {Beeman}, \&
  {Haller}}]{LABOCA}
{Siringo}, G., {et~al.} 2009, \aap, 497, 945

\bibitem[{{Svensson} {et~al.}(2012){Svensson}, {Levan},{Tanvir},
  {Perley},{Michalowski},{Page},{Bloom},{Cenko},
  {Hjorth},{Jakobsson}, {Watson}, \& {Wheatley}}]{slt+12}
{Svensson}, K.~M., {et~al.} 2012, \mnras, 421, 25

\bibitem[{{Tanvir} {et~al.}(2004){Tanvir}, {Barnard},{Blain},
  {Fruchter},{Kouveliotou},{Natarajan},{Ramirez-Ruiz},{Rol},
  {Smith},{Tilanus}, \& {Wijers}}]{tbb+04}
{Tanvir}, N.~R., {et~al.} 2004, \mnras, 352, 1073

\bibitem[{{Tanvir} {et~al.}(2009){Tanvir}, {Fox}, {Levan}, {Berger},
  {Wiersema}, {Fynbo}, {Cucchiara}, {Kr{\"u}hler}, {Gehrels}, {Bloom},
  {Greiner}, {Evans}, {Rol}, {Olivares}, {Hjorth}, {Jakobsson}, {Farihi},
  {Willingale}, {Starling}, {Cenko}, {Perley}, {Maund}, {Duke}, {Wijers},
  {Adamson}, {Allan}, {Bremer}, {Burrows}, {Castro-Tirado}, {Cavanagh}, {de
  Ugarte Postigo}, {Dopita}, {Fatkhullin}, {Fruchter}, {Foley}, {Gorosabel},
  {Kennea}, {Kerr}, {Klose}, {Krimm}, {Komarova}, {Kulkarni}, {Moskvitin},
  {Mundell}, {Naylor}, {Page}, {Penprase}, {Perri}, {Podsiadlowski}, {Roth},
  {Rutledge}, {Sakamoto}, {Schady}, {Schmidt}, {Soderberg}, {Sollerman},
  {Stephens}, {Stratta}, {Ukwatta}, {Watson}, {Westra}, {Wold}, \&
  {Wolf}}]{tfl+09}
---. 2009, \nat, 461, 1254

\bibitem[{{Tanvir} {et~al.}(2011){Tanvir}, {Wiersema}, {Melandri}, {Kruehler},
  {Fynbo}, {Levan}, {Mundell}, {Smith}, {Guidorzi}, {Xu}, \&
  {Skillen}}]{gcn12696}
---. 2011, GRB Coordinates Network, 12696, 1

\bibitem[{{Usui}(2011)}]{gcn12685}
{Usui}, R.~e.~a. 2011, GRB Coordinates Network, 12685, 1

\bibitem[{{van der Horst} {et~al.}(2009){van der Horst}, {Kouveliotou},
  {Gehrels}, {Rol}, {Wijers}, {Cannizzo}, {Racusin}, \& {Burrows}}]{hkg+09}
{van der Horst}, A.~J., {Kouveliotou}, C., {Gehrels}, N., {Rol}, E., {Wijers},
  R.~A.~M.~J., {Cannizzo}, J.~K., {Racusin}, J., \& {Burrows}, D.~N. 2009,
  \apj, 699, 1087

\bibitem[{{Vink} \& {de Koter}(2005)}]{vk05}
{Vink}, J.~S., \& {de Koter}, A. 2005, \aap, 442, 587

\bibitem[{{Wainwright} {et~al.}(2007){Wainwright}, {Berger}, \&
  {Penprase}}]{wbp+07}
{Wainwright}, C., {Berger}, E., \& {Penprase}, B.~E. 2007, \apj, 657, 367

\bibitem[{{Woosley} \& {Bloom}(2006)}]{WoosleyBloom06}
{Woosley}, S.~E., \& {Bloom}, J.~S. 2006, \araa, 44, 507

\bibitem[{{Xin}(2011)}]{gcn12682}
{Xin}, L.~P. e.~a. 2011, GRB Coordinates Network, 12682, 1

\bibitem[{{Xu}(2011)}]{gcn12683}
{Xu}, D.~e.~a. 2011, GRB Coordinates Network, 12683, 1

\bibitem[{{Zhang} {et~al.}(2012){Zhang}, {Burrows}, {Zhang}, {Meszaros},
  {Wang},{Stratta}, {D'Elia}, {Frederiks}, {Golenetskii}, {Cummings}, {Norris},
  {Falcone}, {Barthelmy}, \& {Gehrels}}]{zhang+11}
{Zhang}, B.-B., {et~al.} 2012, \apj, 748, 132

\end{thebibliography}

\clearpage
\begin{deluxetable}{lcccccc|cccc}
\tablecolumns{11} 
\tabcolsep0.03in\footnotesize
\tablewidth{0pt} 
\tablecaption{GROND Observations of GRB\,110709B \label{tab:grond}}
\tablehead{ 
\colhead{UT Date}                &
\colhead{$\delta t$}             &
\colhead{Timespan}               &
\multicolumn{4}{c}{Upper Limits} &
\colhead{Timespan}               &
\multicolumn{3}{c}{Upper Limits} \\
\colhead{}                       &
\colhead{(d)}                    & 
\colhead{(s)}                    & 
\colhead{}                       &
\colhead{}                       &
\colhead{}                       &       
\colhead{}                       &
\colhead{(s)}                    &
\colhead{}                       &
\colhead{}                       &
\colhead{}
}
\startdata
                 &       &      & $g\,'$& $r\,'$& $i\,'$& $z\,'$&      & $J$   & $H$   & $K_{\rm s}$ \\\hline
2011 Jul 9.949        & 0.051 & 585  & 20.76 & 20.53 & 20.32 & 20.49 & 621  & 20.03 & 19.02 & 18.79  \\ 
2011 Jul 10.009       & 0.111 & 2311 & 24.26 & 24.57 & 23.83 & 23.61 & 2335 & 20.83 & 20.00 & 19.27  \\ 
2011 Jul 10.045       & 0.147 & 1633 & 22.61 & 22.80 & 22.39 & 21.12 & 1640 & 19.83 & 18.83 & 18.66  \\\hline
2011 Jul 10.003$\,^a$ & 0.105 & 5275 & 23.63 & 23.89 & 23.28 & 23.09 & 5282 & 21.16 & 19.83 & 19.40
\enddata
\tablecomments{Date specifies the mid-time of the observations, and
timespan is the duration of the observations.  All magnitudes are
upper limits in the AB system, uncorrected for Galactic extinction.\\
$^a$ Results from stacking all three epochs.}
\end{deluxetable}

\clearpage
\begin{deluxetable}{lccl}
\tablecolumns{4}
\tablewidth{0pt}
\tablecaption{JVLA Observations of GRB\,110709B \label{tab:evla1}}
\tablehead{
\colhead{UT Date}    &
\colhead{$\delta t$} &
\colhead{$\nu$}      &
\colhead{$F_{\nu}$}  \\
\colhead{}           &
\colhead{(d)}        &
\colhead{(GHz)}      &
\colhead{($\mu$Jy)}
}
\startdata
2011 Jul 11.97 &  2.07 & 5.8  & $190\pm 14$  \\
2011 Jul 12.92 &  3.02 & 5.8  & $250\pm 27$  \\
2011 Jul 16.88 &  6.98 & 5.8  & $310\pm 18$  \\
2011 Jul 21.97 & 12.07 & 5.8  & $210\pm 14$  \\
2011 Jul 29.96 & 20.06 & 5.8  & $98\pm 17$   \\
2011 Aug 19.85 & 40.59 & 5.8  & $98\pm 6$    \\
2011 Sep 17.70 & 69.72 & 5.8  & $<38\,^a$    \\\hline
2011 Jul 22.01 & 12.11 & 21.8 & $<135$      
\enddata
\tablecomments{Dates specify the mid-time of the observations. \\ $^a$
Upper limits are $3\sigma$.}
\end{deluxetable}

\clearpage
\begin{deluxetable}{lllcl}
\tablecolumns{5}
\tablewidth{0pt}
\tablecaption{Radio Observations of GRB\,111215A \label{tab:evla2}}
\tablehead{
\colhead{Telescope}  &
\colhead{UT Date}    &
\colhead{$\delta t$} &
\colhead{$\nu$}      &
\colhead{$F_{\nu}$}  \\
\colhead{}           &
\colhead{}           &
\colhead{(d)}        &
\colhead{(GHz)}      &
\colhead{($\mu$Jy)}
}
\startdata
JVLA      & 2011 Dec 17.00 & 1.41   &  4.9 & $<115\,^a$  \\
          &                &        &  6.7 & $<93$       \\
          & 2011 Dec 23.05 & 7.46   &  4.9 & $470\pm28$  \\ 
          &                &        &  6.7 & $460\pm34$  \\
          & 2011 Dec 29.03 & 13.44  &  4.9 & $300\pm21$  \\
          &                &        &  6.7 & $540\pm19$  \\
          & 2011 Dec 31.00 & 15.41  &  4.9 & $820\pm30$  \\
          &                &        &  6.7 & $1180\pm30$ \\
          & 2012 Jan 10.99 & 26.40  &  4.9 & $590\pm33$  \\
          &                &        &  6.7 & $820\pm36$  \\
          & 2012 Jan 29.84 & 45.25  &  4.9 & $420\pm52$  \\
          &                &        &  6.7 & $400\pm55$  \\
          & 2012 Mar 12.82 & 88.23  &  4.9 & $330\pm39$  \\
          &                &        &  6.7 & $300\pm34$  \\\hline
          & 2011 Dec 31.00 & 15.41  &  8.4 & $1340\pm63$ \\
          & 2012 Jan 10.99 & 26.40  &  8.4 & $880\pm120$ \\ 
          & 2012 Jan 29.84 & 45.25  &  8.4 & $470\pm81$  \\
          & 2012 Mar 12.82 & 88.23  &  8.4 & $240\pm95$  \\\hline   
          & 2011 Dec 18.93 & 3.35   & 19.1 & $1520\pm74$ \\
          &                &        & 24.4 & $1990\pm47$ \\
          & 2011 Dec 27.08 & 11.49  & 19.1 & $1670\pm47$ \\ 
          &                &        & 24.4 & $1650\pm74$ \\
          & 2011 Dec 31.00 & 15.41  & 19.1 & $1530\pm49$ \\
          &                &        & 24.4 & $1370\pm62$ \\
          & 2012 Jan 10.99 & 26.40  & 19.1 & $840\pm65$  \\   
          &                &        & 24.4 & $730\pm78$  \\
          & 2012 Jan 29.84 & 45.25  & 19.1 & $500\pm66$  \\  
          &                &        & 24.4 & $460\pm69$  \\
          & 2012 Mar 12.82 & 88.23  & 19.1 & $214\pm66$  \\
          &                &        & 24.4 & $180\pm72$  \\\hline
CARMA     & 2011 Dec 17.05 & 1.46   &  93  & $3400\pm290$ \\
          & 2011 Dec 20.02 & 4.43   &  93  & $2500\pm330$ \\ 
          & 2011 Dec 22.04 & 6.45   &  93  & $2500\pm300$ \\
          & 2011 Dec 24.10 & 8.51   &  93  & $2200\pm270$ \\   
          & 2011 Dec 26.01 & 10.42  &  93  & $2200\pm270$ \\
          & 2012 Jan 1.14  & 16.55  &  93  & $1500\pm340$ \\\hline  
SMA       & 2011 Dec 18.30 & 2.71   &  230 & $<2600$  
\enddata
\tablecomments{Dates specify the mid-time of the observations. \\ $^a$
Upper limits are $3\sigma$.}
\end{deluxetable}

\clearpage
\begin{deluxetable}{lcccc}
\tablecolumns{5} 
\tabcolsep0.1in\footnotesize
\tablewidth{0pt} 
\tablecaption{Results of Broad-band Afterglow Modeling for GRB\,110709B
\label{tab:110709model}}
\tablehead{
\colhead{Parameter}   &
\colhead{$z=1$}       &
\colhead{$z=2$}       &
\colhead{$z=3$}       &       
\colhead{$z=4$}            
}
\startdata
$E_{\rm K,iso,52}$ (erg)       & 0.5   & 2.8   & 7.1   & 13.2  \\
$A^*$ ($5\times 10^{11}$ g cm$^{-1}$) & 4.0   & 5.7   & 7.2   & 8.5   \\
$\bar{\epsilon}_e$                    & 0.18  & 0.11  & 0.08  & 0.07  \\
$\epsilon_B$                          & 0.002 & 0.002 & 0.002 & 0.002 \\
$p$                                   & 2.07  & 2.06  & 2.05  & 2.05  \\
$t_j$ (d)                             & 3.65  & 3.45  & 3.27  & 3.11  \\
$\theta_j (rad)$                      & 0.40  & 0.25  & 0.19  & 0.16  \\
$E_{\rm K,51}$ (erg)                  & 0.4   & 0.9   & 1.3   & 1.7   \\
$E_{\rm \gamma,iso,52}\,^a$ (erg)     & 7.0   & 26   & 52   & 83   \\
$E_{\rm \gamma,51}$ (erg)             & 5.4   & 8.0   & 9.4   & 10.6  \\
$A_V\,^{b}$ (mag)                     & $\gtrsim 10.5$ & $\gtrsim 5.3$ & $\gtrsim 4.4$ & $\gtrsim 3.4$   
\enddata
\tablecomments{See \S 3 for description of model parameters. \\
$^a$ Using the Konus-WIND $20-5000$ keV fluence. \\
$^b$ Using the SMC extinction curve.}
\end{deluxetable}

\clearpage
\begin{deluxetable}{lcc}
\tablecolumns{2} 
\tabcolsep0.1in\footnotesize
\tablewidth{0pt} 
\tablecaption{Results of Broad-band Afterglow Modeling for GRB\,111215A 
\label{tab:111215model}}
\tablehead{
\colhead{Parameter} & 
\colhead{$z=2$}     &           
\colhead{$z=3$}              
}
\startdata
$E_{\rm K,iso,52}$ (erg)       & 4.3   & 10.6 \\
$A^*$ ($5\times 10^{11}$ g cm$^{-1}$) & 19    & 26   \\
$\bar{\epsilon}_e$                    & 0.22  & 0.18 \\
$\epsilon_B$                          & $2.2\times 10^{-4}$ & $1.5\times 10^{-4}$\\
$p$                                   & 2.35  & 2.35 \\
$t_j$ (d)                             & 12    & 12   \\
$\theta_j$ (rad)                      & 0.42  & 0.33 \\
$E_{\rm K,51}$ (erg)                  & 3.7   & 5.7  \\
$E_{\rm \gamma,iso,52}\,^a$ (erg)     & 4.5  & 9.0 \\
$E_{\rm \gamma,51}$ (erg)             & 3.9   & 4.9  \\
$A_V\,^{b}$ (mag)                     & $\gtrsim 8.5$ & $\gtrsim 6.8$  
\enddata
\tablecomments{See \S 3 for description of model parameters. \\
$^a$ Using the fluence in the {\it Swift}/BAT $15-150$ keV band. \\
$^b$ Using the SMC extinction curve.}
\end{deluxetable}

\clearpage

\clearpage

\clearpage

\clearpage

\clearpage

\clearpage

\clearpage

\clearpage

\end{document}